\documentclass[pre,aps,floatfix,letterpaper,reprint]{revtex4-1}

\usepackage{graphicx}
\usepackage{amsmath}
\usepackage{float}
\usepackage{placeins}
\usepackage{bm}
\usepackage{xcolor}
\usepackage{ulem}

\definecolor{green}{RGB}{0,100,0}

\newcommand{\blue}{\color{black}}

\begin{document}
\title{Mechanisms leading to axial segregation of bidisperse particles in rotating tumblers}

\author{Umberto D'Ortona}
\email{umberto.dortona@cnrs.fr}
\affiliation{Aix-Marseille Univ., CNRS, Centrale Med., M2P2, Marseille, France}
\author{Richard M. Lueptow}
\affiliation{Department of Mechanical Engineering, 
         Northwestern University, Evanston, Illinois 60208, USA}
\affiliation{Department of Chemical and Biological Engineering, Northwestern University, Evanston, Illinois 60208, USA}
\author{Nathalie Thomas} 
\affiliation{Aix-Marseille Univ., CNRS, IUSTI, Marseille, France}

\date{\today}
\begin{abstract}
The origin of segregated axial bands for size-bidisperse particles in long rotating tumblers is a long-standing question in granular flows and a problem of both fundamental and practical interest. Using DEM simulations, we consider {\blue the streamwise and time-dependent development of the underlying instability as well as performing a broad parametric study over} a wide range of particle size ratios, tumbler lengths and diameters, species fractions, and fill levels to determine the conditions under which axially segregated bands appear. We show that initial radial segregation of the small and large particles in the surface flowing layer results in a layer of mixed particles with a higher bulk density due to more efficient packing over a layer of nearly monodisperse small particles with a lower bulk density. This leads to a granular Rayleigh-Taylor-like instability that produces recirculation cells driving axial band formation. The appearance of a layer of high density mixed particles over a layer of lower density monodisperse particles can be directly connected to the operating conditions, explaining why axially segregated bands appear only under certain conditions. Typically, increasing the particle size ratio improves packing and increases the density of the mixed layer, enhancing the Rayleigh–Taylor instability and accelerating axial segregation. 
\end{abstract}
\maketitle

\section{Introduction}

Partially filled horizontal cylindrical rotating tumblers, or drums, are frequently used as a canonical
system for studying the flow and segregation of granular materials \cite{OttinoKhakhar00, MeierLueptow07, Ottino06} because similar systems are commonly used in industrial processes for mixing,
coating, and granulation. For an
initially well-mixed bidisperse system of particles of two sizes, the smaller particles fall through interstices between large particles to
lower levels in the thin surface flowing layer before they deposit onto the fixed bed of particles in solid body rotation with the tumbler. The result after a few rotations is a radial segregation pattern, when viewed in cross-section, with large particles at the periphery of the bed of particles and small particles at the core  \cite{ClementRajchenbach95, CantelaubeBideau95, DuryRistow1997, KhakharMcCarthy97}.

\begin{figure}[hptb] 
\includegraphics[width=0.9\linewidth]{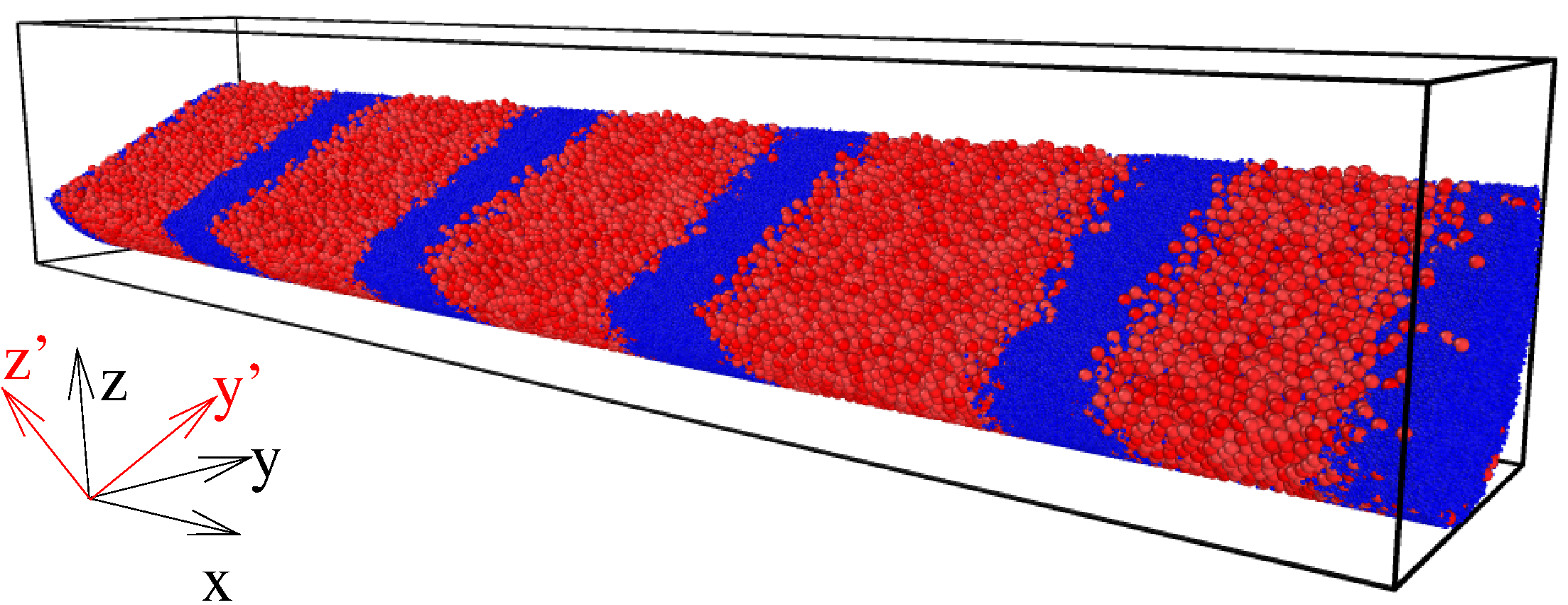}
\caption{Example of steady-state axial segregation at about 1400~s (350 rotations) in a tumbler that is 1~m long with periodic boundary conditions at each end, 18~cm in diameter, rotating at 15~rpm, and filled to 50\% by volume with equal volume fractions of 2~mm (blue) and 6~mm (red) spherical particles.}
\label{d3w500}
\end{figure}

Less well understood than radial segregation until recently is the subsequent formation of axially spaced bands rich in small or large particles that appear at the free surface and tumbler wall along the length of a long tumbler; an example is shown in Fig.~\ref{d3w500}.  
Although first reported in 1940 by Oyama~\cite{Oyama40}, axial segregation did not attract substantial attention until much later \cite{DonaldRoseman62, DasguptaKhakhar91, Nakagawa94, ZikLevine94, HillKakalios94, HillKakalios95, ChooMorris98, Levine99, Rapaport02, AlexanderMuzzio04}. The axial bands typically have a wavelength of approximately one tumbler diameter~\cite{CharlesMorris06, FiedorOttino06, JuarezLueptow08, JuarezOttino10}, although the bands eventually tend to merge and coarsen \cite{Frette97, HillCaprihan97, ChooMorris97, ChooMorris98, FiedorOttino03, FiedorOttino06, JuarezLueptow08, JuarezOttino10}. The bands occur under a wide range of conditions, including various rotation speeds, fill levels, particle sizes and shapes, and species concentrations, as well as in tumblers with square or circular cross-sections, with smooth or rough walls \cite{DOrtonaThomas16,DOrtonaThomas17}, in rotating spheres \cite{ChenLochman09,DOrtonaThomas16,DOrtonaThomas17}, with submerged as well as dry particles, and in tilted tumblers \cite{JainKhakhar01,FiedorOttino03, Arndt05, JuarezLueptow08}. When observed using an index-matched interstitial fluid or with magnetic resonance imaging, what appears as a band of small particles
is actually a region where the core of small particles reaches the flowing layer surface and tumbler wall; bands of large particles are regions where an annulus of larger particles surrounds a small radius core of small particles \cite{HillKakalios95, JainKhakhar01, KhanMorris2004, Arndt05, TaberletNewey06, ChenLueptow11}.

Proposed mechanisms for band formation have been based on differing angles of repose for small and large particles ~\cite{KhanMorris2004, ZikLevine94, HillKakalios94, HillKakalios95}, a concept called "negative diffusivity"~\cite{HillKakalios94, HillKakalios95, Aranson1999a, Aranson1999b}, differing mobility of particle species~\cite{ZikLevine94, Levine99}, bands of large particles forming near frictional endwalls~\cite{ChenLueptow11}, and a capillary-like instability based on subsurface bulges of the radially segregated small particle core reaching the free surface~\cite{AlexanderMuzzio04}.  However, none of these mechanisms can explain the source of the initial subsurface bulges or the origin of bands far from endwalls or with periodic boundary conditions replacing endwalls.  

In contrast, {\blue our recent letter} demonstrated~\cite{DOrtonaLueptow24} that the mechanism for axial band formation is a recently discovered granular flow instability~\cite{DOrtonaThomas20,DOrtonaMartinand25} similar to the fluid Rayleigh-Taylor (RT) instability,
which occurs when a dense fluid lies atop a less dense fluid. Unlike the fluid instability, where the two fluid layers are static, the granular layers must be flowing. A layer of large, dense particles above a layer of small, light particles flowing down an inclined plane destabilizes to form upward plumes of the small, light particles and downward plumes of large, dense particles~\cite{DOrtonaThomas20}. These plumes appear at the free surface as parallel stripes extending in the streamwise direction. The velocity field in a cross-section of the flow consists of counter-rotating rolls between the plumes. Even when starting with initially mixed particles, the particles first segregate to form a thin surface layer of large, dense particles that then destabilizes such that the entire flow organizes into streamwise convective rolls and surface bands of the two particle species~\cite{DOrtonaThomas20}.

{\blue The unique aspect of the formation of axial bands in a rotating tumbler  is that, although it arises from the same RT instability as for particles varying in both size and density~\cite{DOrtonaThomas20,DOrtonaMartinand25}, the mechanism for axial segregation occurs for size-bidisperse particles with the \textit{same} mass density~\cite{DOrtonaLueptow24}  -- no density difference between the two species of particles is needed.  Instead, the} excess density in the upper layer necessary for the RT instability arises from a unique property of granular materials: a mixture of grains of different sizes has a higher packing fraction, and thus a higher effective density, than monodisperse particles~\cite{Pilliteri20}. 
It is this {\blue higher packing density in the} mixed-particle interfacial layer between the upper monodisperse layer of large particles and the lower monodisperse layer of small particles that induces the RT instability and, eventually, the formation of bands~\cite{DOrtonaLueptow24}{\blue , not a density difference due to the particle densities themselves as was the case when the granular RT instability was first identified~\cite{DOrtonaThomas20,DOrtonaMartinand25}. The packing density driven RT instability also} occurs when the size-bidisperse mixtures of particles flow down an incline, leading to longitudinal bands of small and large particles~\cite{DOrtonaLueptow26}. {\blue The unique situation in a rotating tumbler is that the particles continually enter and exit the flowing layer leading to dependence on geometric and time-dependent parameters inherent in rotating tumblers.}

This paper expands our recent {\blue letter}~\cite{DOrtonaLueptow24}, in which we {\blue briefly} introduced the granular RT instability as the mechanism for axial band formation and examined the influence of the particle size ratio on band formation. Section II presents the simulation methods used. Section III discusses {\blue the development of the RT instability along the length of flowing layer and with time to understand the mechanisms responsible for the eventual band formation associated with axial segregation.} Section IV provides an extensive study of the influence of broad range of tumbler and particle parameters {\blue including tumbler length and diameter,
species fractions, and fill levels to determine the conditions under which axially segregated bands appear.  We also explore the limits of band formation as the size ratio becomes small and how small changes in the density ratio can accelerate band formation or suppress it altogether.  We do not consider here} rotation speed, gravity, and tumbler diameter (while keeping the particle size ratio constant){\blue , whose combined effects are often expressed in terms of a Froude number. This} will be addressed in a {\blue forthcoming} companion article.

\section{Numerical model}
\label{numericalmodel}

Band formation in partially filled cylindrical tumblers is explored using DEM simulations~\cite{CundallStrack79}, specifically the open-source DEM code LIGGGHTS~\cite{Kloss2012}. 
The cylindrical tumbler wall is modeled as a smooth frictional surface 
with infinite mass for calculating the collision force between the
tumbling particles and the wall. To avoid endwall friction effects, the two ends of the cylindrical tumbler section that is simulated utilize periodic boundary conditions so that particles leaving the right side re-enter on the left side and vice versa.

The horizontal cylindrical tumblers considered here typically have an inner diameter $D=2R=0.18$~m, corresponding to $D/d=90$, where $d=2$~mm is the small particle diameter. The tumbler lengths range from $L=0.1$~m to $L=1$~m, depending on the particular aspect of the RT instability and band formation being considered. The $x$-axis is the axis of 
rotation, the $z$-axis is opposite to $\bm g$, and the $y$-axis is perpendicular to $x$ and $z$. The origin of the coordinate system is on the axis of rotation at the left end of the tumbler. In the analysis of the results, we use a second coordinate system, ($x,\,y',\,z')$, where the $y'$-axis is parallel to the free surface, and the $z'$-axis is perpendicular to the free surface (see Fig.~\ref{d3w500}).

Unless otherwise stated, tumblers are filled to volume fractions (fill levels) of
$V=0.5$  with equal volumes of small and large particles such that the average particle species volume fraction in the tumbler is $F_s=F_l=0.5$, where subscripts $s$ and $l$ refer to small and large particles, respectively. The large particle diameter, $d_l>2$~mm, is adjusted to vary the particle size ratio, $d_l/d$. We also consider small particle sizes ranging from $d=1$~mm to $d=2.9$~mm. 
The particle properties correspond to cellulose
acetate: density $\rho =$ 1308~kg~m$^{-3}$, restitution coefficient $e = 0.87$ \cite{DrakeShreve86,FoersterLouge94,SchaferDippel96}. The effect of the density ratio is considered by varying the density of the large particles in the range $0.87\rho \leq \rho_l \leq 1.05\rho$. The total number of particles 
ranges from about $9 \times 10^4$ to $1 \times 10^6$. 
To avoid a close-packed ordered structure, the particles have a
uniform size distribution ranging from 0.95$d$ to 1.05$d$ for small particles and from 0.95$d_l$ to 1.05$d_l$ for large ones. 
The friction
coefficient between particles and between particles and the walls is set to 
$\mu = 0.7$.  
Gravitational acceleration is $g = 9.81$~m~s$^{-2}$. Particle interactions are calculated using a linear contact model with Young's modulus $E=5\times 10^7$ Pa and Poisson's ratio $\nu=0.3$. A standard tangential force model with Coulomb friction and elasticity is implemented~\cite{CundallStrack79,Kloss2012}.
The integration time step is typically $\delta t =5\times 10^{-6}$~s for numerical stability, consistent with previous simulations \cite{TaberletNewey06,ChenLueptow11,ZamanDOrtona13}, and is reduced progressively to $\delta t =2.5\times 10^{-6}$~s for $d=1$~mm particles.   
Note that a larger time step results in much stronger and faster axial segregation that is likely nonphysical, an effect that appears to be related to relative slip between particles. {\blue Typically, for 2 and 4 mm particles, increasing the time step to $\delta t = 6 \times 10^{-6}$ does not affect axial segregation. However, for larger time steps, axial segregation progressively accelerates and the stationary level of axial segregation becomes more pronounced. For $\delta t = 1 \times 10^{-5}$, axial segregation develops approximately four times faster. More critically, at this time step, axial segregation occurs in tumblers with a diameter of $D = 60 d$, where no axial segregation is expected.
This is discussed in detail in the Supplementary Material (SM) \cite{SM}.}

The particles are initially distributed randomly throughout the entire tumbler volume with zero gravity. Then gravity and tumbler rotation are turned on simultaneously so that the particles fall to the
bottom of the tumbler and begin to flow due to tumbler rotation. 
The tumbler typically rotates at $\omega=15$~rpm $=1.57$~rad/s, corresponding to a Froude number
${\rm Fr}=\omega^2 R/g= 0.023$, although other tumbler radii, $R=0.06$ to  0.16~m ($0.015\le {\rm Fr}\le 0.04$), are also considered. 
This range of Fr corresponds
to the continuous-flow rolling regime for tumbler flow, characterized by a
steady flowing layer with a surface that is essentially flat, although it may
be slightly S-shaped for larger values of ${\rm Fr}$~\cite{MeierLueptow07}. 
The position and velocity of each particle are stored every 0.1~s for general postprocessing, and every 0.0005~s for generating the velocity field when the flow and segregation are stationary.

One of our objectives is to find the conditions for which the system is most unstable, that is, the conditions for which the small and large particle bands form most quickly and are maximally segregated. To quantify the evolution of axial segregation, we measure an axial segregation index $I_\mathrm{seg}(t)=(1/L)\int_0^L |F_l-f_l(x,t)|+|F_s-f_s(x,t)|{\rm d}x$, where $f_i(x,t)$ is the volume fraction for species $i$ averaged across the $y-z$~plane in the particle bed at axial position $x$ and time $t$, and $F_i$ is the overall volume fraction for species $i$, noting that $f_l+f_s=1$ locally and $F_l+F_s=1$ globally. 
$I_\mathrm{seg}=0$ corresponds to no axial segregation, which can occur either because the two particle species are completely mixed or because there is an axially invariant radial segregation comprising a radial core of small particles surrounded by a periphery of large particles at the free surface and tumbler wall. For perfect axial segregation consisting of axial bands of pure small or pure large particles, $I_\mathrm{seg}=1$ for $F_l=F_s=0.5$. Note that if $F_l \neq F_s$, a prefactor of $(4F_l F_s)^{-1}$ is necessary in the definition of the segregation index so that $I_{\rm seg}$ remains in the range $[0,\, 1]$. 
The time evolution of $I_\mathrm{seg}$ indicates how quickly axial segregation occurs, and the value at which $I_\mathrm{seg}$ plateaus indicates the maximum, steady-state degree of segregation. Note that in many experiments on axial segregation, it is only the appearance of bands at the surface or through clear tumbler walls that can be observed~\cite{Frette97, ChooMorris97, FiedorOttino03, JuarezLueptow08}, even though axial segregation can also manifest as subsurface bulges in the radial segregation core~\cite{HillKakalios95, JainKhakhar01, Rapaport02, KhanMorris2004, Arndt05, TaberletNewey06, ChenLueptow11} that are not usually visible. Both these subsurface bulges in the radial segregation and the visually observable bands are {\blue captured by $I_\mathrm{seg}$. Hence, not only does $I_\mathrm{seg}$ detect subsurface bulges that eventually lead to segregated bands of particles, it avoids the qualitative subjectivity of basing the appearance of surface bands on when they are visible at the surface, which varies depending on the contrast in the colors of the two particles species, the particle sizes, and the tumbler diameter, as well as avoiding the subjectivity in determining what actually constitutes a particle band at the free surface (a few particles of the smaller species appearing at the free surface versus a monodisperse small particle band that is several particle diameters wide).}

Critical to understanding the RT instability is the local effective density of the medium, which is smaller for monodisperse large or small particles and larger for the bidisperse packing of a mixture of large and small particles. Hence, we measure the local volume fraction, or compacity, of all particles, $\phi$. Often a binning method is used to measure $\phi$, but this is difficult here because the large particles may be larger than the bin size needed for adequate spatial resolution. Instead, a subgrid measurement with spacing $\Delta x=d/6$ is used, checking at each node of the grid whether it is inside a particle or
not, and if so, whether it is inside a small or a large particle. Averaging over $6^3$ nodes
results in a smoothed measure of $\phi$. The same subgrid method is used to compute the velocity field. A period of 100~s of averaging under stationary conditions is necessary to obtain an adequately smooth velocity field for integrating mean particle trajectories.

Rayleigh-Taylor instabilities are often studied by considering the growth rate of an initial sinusoidal perturbation of the interface between layers having different densities. Using a similar approach here, a sinusoidal modulation of the particle species fraction is introduced along the length of the tumbler prior to turning on gravity and tumbler rotation. 
Specifically, the periodic tumbler section is divided into 16 axial slices. Large particles are distributed in these slices according to a sinusoidal variation with amplitude $\Delta f_l$ so that the maximum large particle fraction is $F_l+\Delta f_l$ and the minimum is $F_l-\Delta f_l$, where typically $0.1 \leq \Delta f_l \leq 0.3$. Then small particles are added, resulting in a sinusoidal distribution for small particles that is out of phase with that of the large particles. With this initial condition, the particles reach a smooth sinusoidal distribution within one tumbler rotation (4 s),  after which the modulation amplitude can either grow or decay. 
Exponential growth of the modulation amplitude typically precedes band formation, with the vertical position of the interface between radially segregated large and small particles described by an equation of the form 
$z(x,t)=B_0 \exp(\gamma t) \sin(2\pi x/\lambda)$, where $\gamma$ is the growth rate and $\lambda$ is the wavelength \cite{Chandrasekhar61,Charru}. While $\lambda$ is relatively easy to measure, estimating $\gamma$ is more difficult because determining the position of the large-small particle interface is not straightforward, as it undergoes numerous transformations as individual particles make their circuits through the flowing layer and static zone. In addition, the particle velocity in the flowing layer depends on its depth within the flowing layer. Fortunately, the amplitude of the sinusoidal perturbation and the segregation index are proportional and both grow exponentially, allowing us to obtain a measure of the growth rate using $I_{\rm seg}$, as we demonstrate in the SM~\cite{SM}. 

\section{Mechanisms} \label{mechanisms}
We have previously shown~\cite{DOrtonaLueptow24} that axial segregation of size-bidisperse particles in a rotating tumbler results from an RT instability where the excess density arises from a property of granular media: a packing of particles of different sizes is more compact, and therefore denser, than a packing of monodisperse particles. Here we expand on those initial results and consider the mechanisms responsible for the instability in greater detail. 

\subsection{Axial segregation, a mixture-induced RTI}

\begin{figure}[hptb] 
\includegraphics[width=\linewidth]{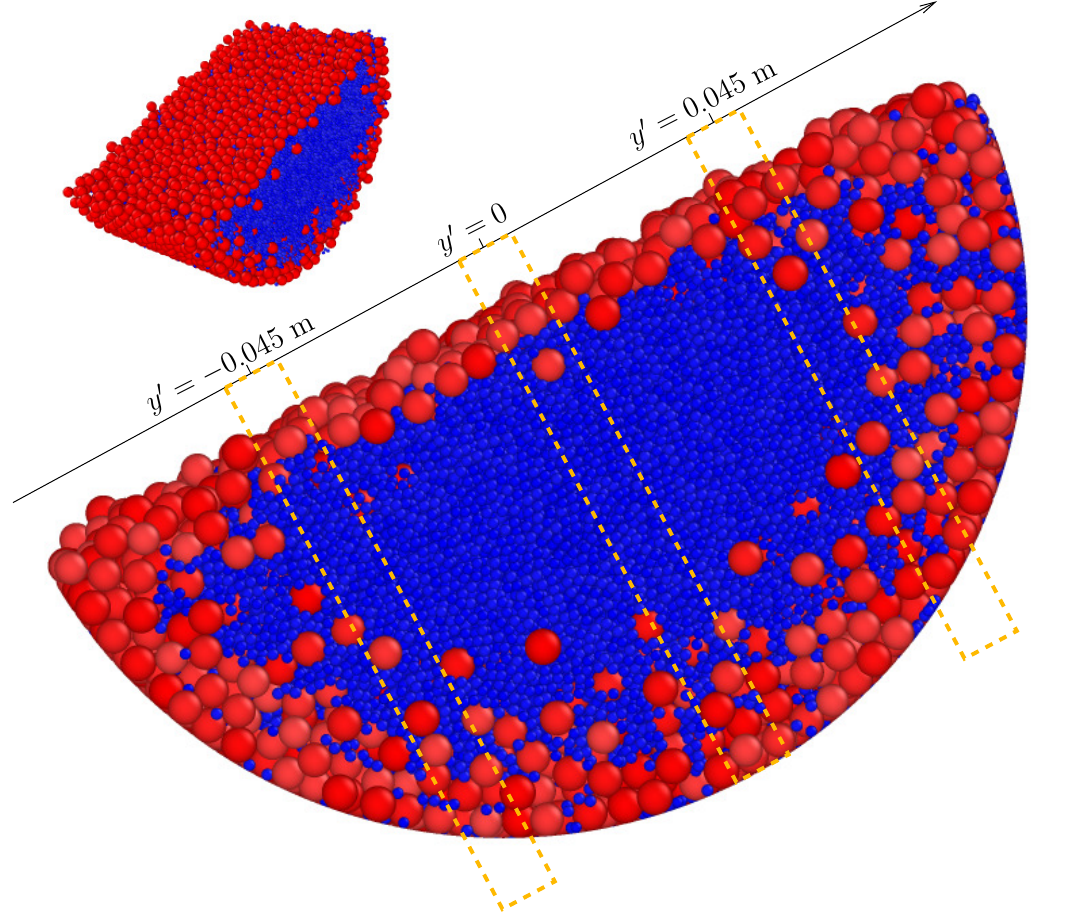}
\caption{Side view of a rotating tumbler filled to volume fraction $V=0.5$ with equal volumes of large ($d_l=6$~mm) and small ($d=2$~mm) particles ($F_l=F_s=0.5$, $d_l/d=3$). The length of the tumbler is $L=50d=10$~cm to prevent axial segregation and the end conditions are periodic. The $D=90d=18$~cm tumbler rotates at $\omega=15$~rpm. The image is at $t=100$~s, after 25 rotations, when a steady-state radial segregation is well established. The inset shows the entire simulated tumbler. Dashed boxes (1~cm wide and centered on $y'=-0.045,\, 0$, and 0.045~m) indicate regions where the volume fraction (compacity) profiles are measured.}
\label{d3side}
\end{figure}

The RT instability requires a density inversion in a gravitational field. To demonstrate the density inversion in a rotating tumbler where both large and small particles have the same density, consider a tumbler with $L=10$ cm, which is so short that axial segregation cannot develop.  After 25 tumbler rotations, a core of small blue particles is surrounded by large red particles at the free surface and the periphery of the bed of particles, as shown in Fig.~\ref{d3side}, which is typical of steady-state radial segregation.

\begin{figure}[hptb] 
\includegraphics[width=\linewidth]{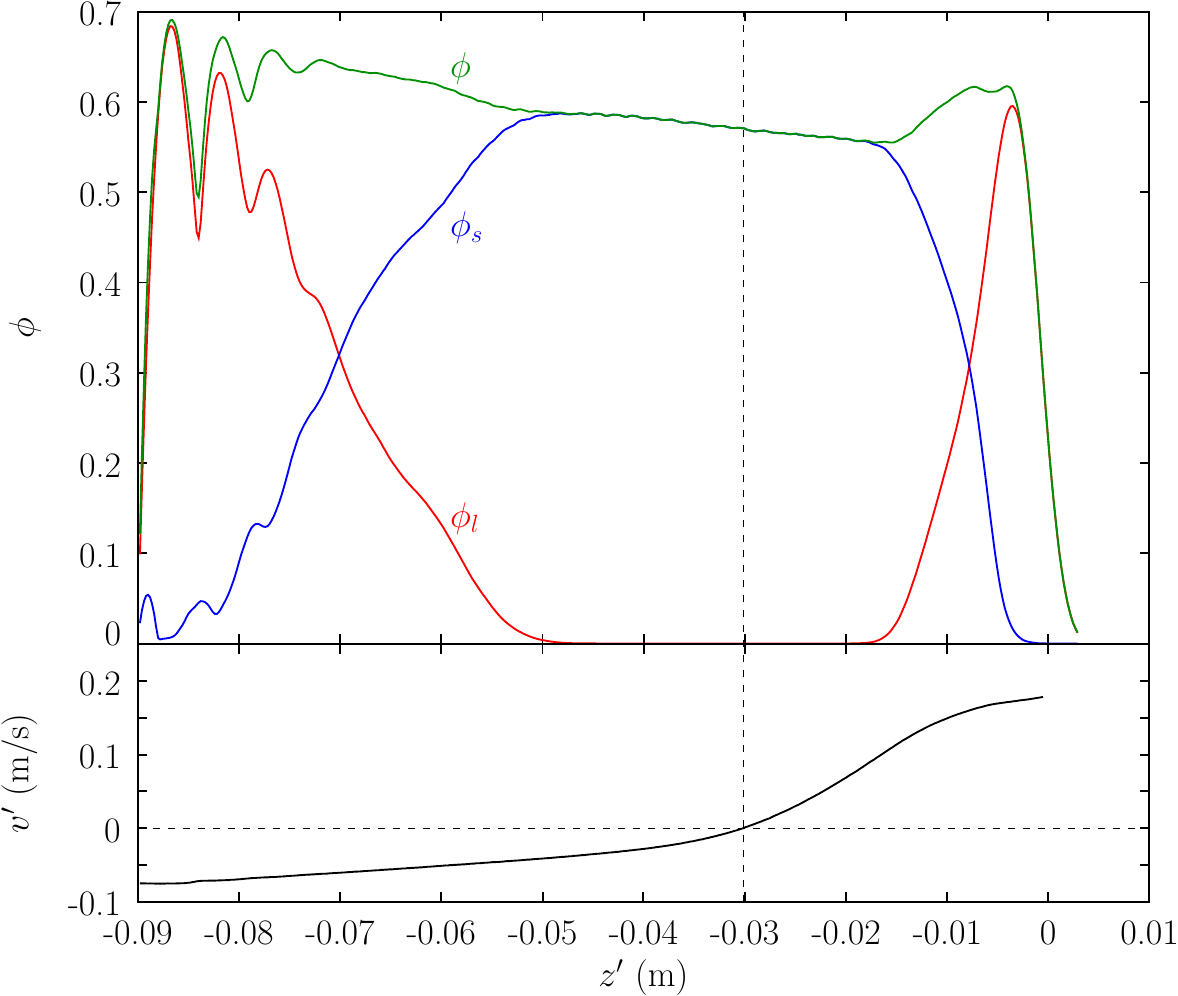}
	\caption{Volume fraction profiles of small ($\phi_s$, blue) and large ($\phi_l$, red) particles, the sum of both ($\phi$, green), and the streamwise velocity profile (black) at $y'=0$ (dashed region in Fig.~\ref{d3side}) averaged from $t=~80$ to 100~s. Positive velocity corresponds to particles are flowing down the incline.}
\label{d3profil}
\end{figure}

Figure~\ref{d3profil} shows the volume fraction (also called compacity or packing fraction, $\phi_i$) and streamwise velocity ($v'$) profiles measured at the center of the tumbler at $y'=0$ (averaged in the $y'-$direction in the dashed orange box in Fig.~\ref{d3side}) {\blue through the entire depth of the particle bed}. Zero streamwise velocity in the laboratory reference frame, $v'=0$, is generally used as the criterion separating the static zone from the flowing layer (vertical dashed line). Ignoring creeping flow that occurs in the static zone just below this point, the remainder of the velocity profile in the static zone is linear, reflecting solid body rotation. Radial segregation is evident in the volume fraction profiles in the static zone ($z'\lesssim -0.03$~m). 

In the flowing layer ($z'\gtrsim -0.03$~m), the velocity increases from zero at the bottom of the flowing layer to its maximum at the free surface. More important for the granular RT instability are the volume fractions of small and large particles.  The volume fraction of small particles, $\phi_s$, reflects pure small particles in the lower part of the flowing layer and decreases to zero moving toward the free surface. The volume fraction of large particles, $\phi_l$, is zero in the lower portion of the flowing layer but increases near the free surface, reflecting the segregated layer of large particles above the layer of monodisperse small particles. A zone of mixed large and small particles is evident as a bump in the total particle volume fraction, $\phi=\phi_l+\phi_s$, for $-0.015 \gtrsim z' \gtrsim 0$~m, {\blue which we previously identified as crucial to the RT instability~\cite{DOrtonaLueptow24}}. The maximum volume fraction of $\phi\simeq 0.62$ occurs at $z'\simeq -0.005$~m, while $\phi\simeq 0.57$ in the region of pure small particles just below it, reflecting a difference of nearly 10\% in the particle volume fraction and, hence, the density. This higher-density mixed-particle layer near the free surface and above the lower density layer of small particles is responsible for the RT instability that leads to axial segregation. 

\begin{figure}[hptb] 
\includegraphics[width=\linewidth]{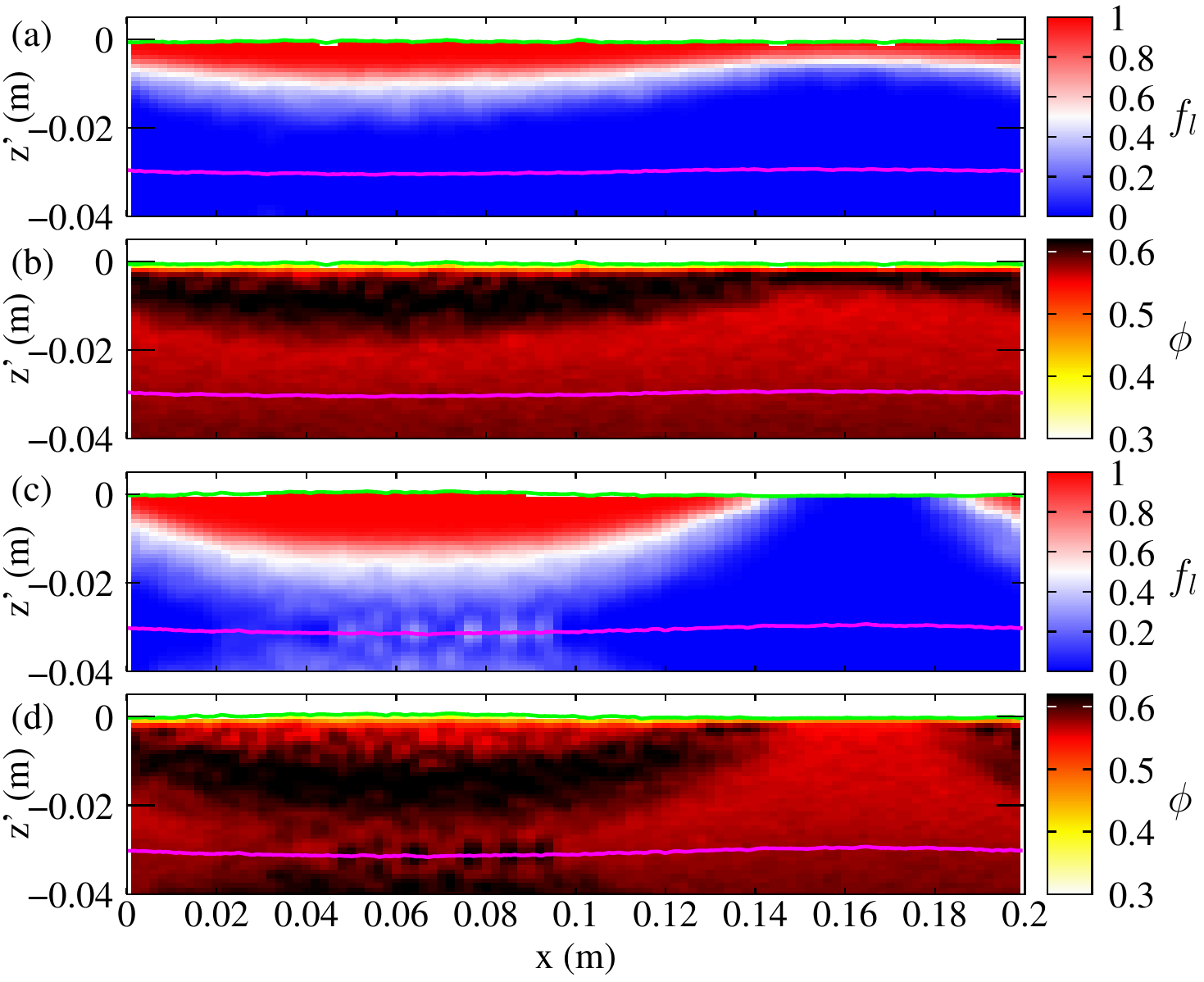}
	\caption{(a, b) Local large particle species fraction, $f_\mathrm{l}$, and total particle volume fraction, $\phi$, where averaging is over $t=50 \pm 10$~s, and (c, d) where averaging is over $t=1000 \pm 10$~s in the $x-z'$~plane at $y'=0$ (dashed region in Fig.~\ref{d3side}) for a periodic cylindrical tumbler with $L=100d$, $D=90d$, $\omega=15$~rpm, and filled to $V=0.5$ with equal volumes ($F_l=F_s=0.5$) of $d=2$~mm and $d_l=6$~mm spherical particles ($d_l/d=3$). The green line indicates the free surface (based on $\phi=0.3$), and the magenta line delineates the flowing layer and static zone (based on $v'=0$).}
\label{color}
\end{figure}

The nature of the granular RT instability resulting from the density inversion is evident in Fig.~\ref{color}, which
shows the local large particle fraction, $f_l$, and volume fraction, $\phi$, fields in an $L=20$~cm long tumbler with periodic end conditions. Unlike the $L=10$~cm tumbler used in Figs.~\ref{d3side} and~\ref{d3profil}, the $L=20$~cm tumbler is long enough for the RT instability to occur, as we show later in the paper. The other parameters
remain unchanged from Figs.~\ref{d3side} and~\ref{d3profil}. Although we have shown previously that a wavy interface can develop naturally from a density inversion in the flowing layer~\cite{DOrtonaLueptow24}, here the instability is initiated by filling the tumbler with sinusoidally modulated species fractions with amplitude $\Delta f_l=0.3$ 
and wavelength $\lambda = L$, such that after radial segregation (occurring in about 5 rotations), the left part of the tumbler is richer in large particles. 

The distributions of large and small particles in the flowing layer around $t=50$~s, shortly after radial segregation, are shown in Fig.~\ref{color}(a). Due to the sinusoidal initial perturbation, a wavy interface between large and small particles (white) occurs in the flowing layer (above the magenta line). Figure~\ref{color}(b) shows the associated volume fraction, $\phi$. The volume fraction is higher in the mixture zone, $\phi\approx 0.62$, than in the zone of pure small particles, where $\phi\approx 0.55$. 
The maximum of $\phi\approx 0.62$ corresponds to $f_l\approx 0.72$ (pink on the color scale). 

Unlike the fluid RT instability, where the system eventually stabilizes after the complete reversal of the dense and light phases, in this case (and also on an incline \cite{DOrtonaLueptow26}) the dense zone becomes thicker in regions with more large particles, and axial segregation appears as a zone of pure small particles with lower $\phi$ emerges at the free surface. Eventually, the system freezes into a stationary state of alternating bands of large and small particles, as shown in Fig.~\ref{color}(c) for $t=1000$~s. An annulus of large particles surrounds a core of small ones at the large-particle band, and a region of small particles extends through the entire bed of particles to the surface at the small-particle band. The associated spatial distribution for $\phi$ is shown in Fig.~\ref{color}(d). The density excess corresponding to the mixed zone has thickened but is now topped by a lower-density zone of large particles at the surface.
Of course, here the dense phase is not a fluid in the classical sense but results from the mixture of two sizes of particles. Hence, the density excess depends on other phenomena, specifically size segregation and collisional diffusion, which in turn depend on the flow and shear, both influenced by the position within the rotating tumbler, as well as particle properties such as the size ratio~\cite{DOrtonaLueptow24}.

\begin{figure*}[htbp] 
\includegraphics[width=\linewidth]{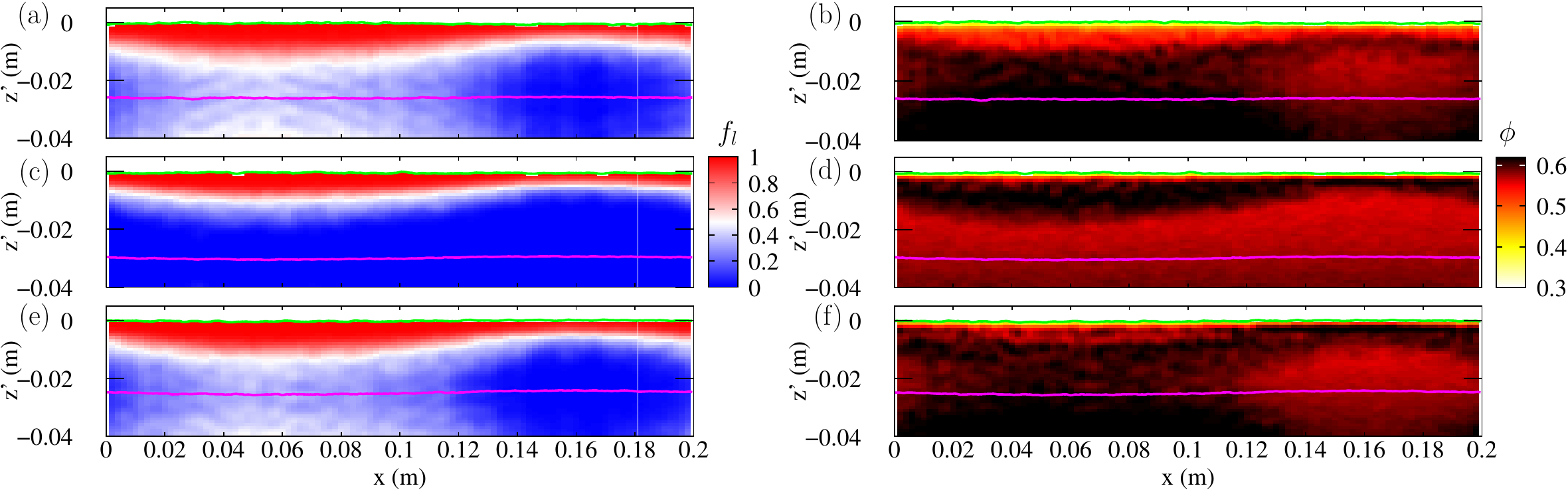}
\caption{Large particle fraction, $f_\mathrm{l}$, and total particle volume fraction, $\phi$, at time $t=50 \pm 10$~s in the $x-z'$~plane at (a, b) $y'=0.045 \pm 0.005$~m (upstream portion of the flow in Fig.~\ref{d3side}), (c, d) $y'=0 \pm 0.005$~m, and (e, f) $y'=-0.045 \pm 0.005$~m (downstream portion of the flow). Same parameters as Fig.~\ref{color}. 
}
\label{colorinit}
\end{figure*}

\begin{figure*}[htbp] 
	\includegraphics[width=\linewidth]{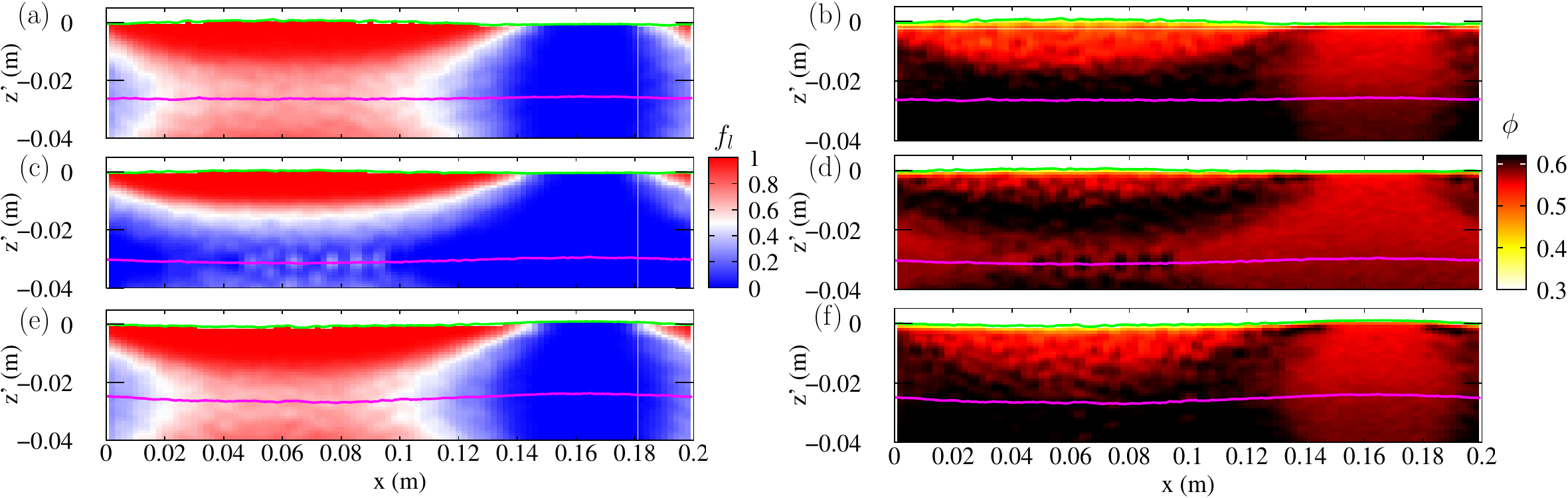}
	\caption{Large particle fraction, $f_\mathrm{l}$, and total particle volume fraction, $\phi$, at time $t=1000 \pm 10$~s in the $x-z'$~plane at (a, b) $y'=0.045 \pm 0.005$~m (upstream portion of the flow in Fig.~\ref{d3side}), (c, d) $y'=0 \pm 0.005$~m, and (e, f) $y'=-0.045 \pm 0.005$~m (downstream portion of the flow). Same parameters as Fig.~\ref{color}.}
	\label{colorstatio}
\end{figure*}

{\blue Although our initial letter~\cite{DOrtonaLueptow24} identified the RT instability as the mechanism to initiate band formation, to} better understand the development of the granular RT instability along the flow direction, $f_l$ and $\phi$ are measured in the upstream portion ($y'=0.045 \pm 0.005$~m) and the downstream portion ($y'=-0.045 \pm 0.005$~m) of the flowing layer at $t=50$~s and $t=1000$~s, corresponding to Figs.~\ref{colorinit} and \ref{colorstatio}, respectively.
To facilitate the comparison, $f_l$ and $\phi$ at $y'=0$ are also included in the figures. 
Early in the development of the RT instability, the $f_l$ fields in the upstream, central, and downstream parts of the flowing layer [Fig.~\ref{colorinit}(a, c, e)] are similar, except that the thickness of the large particle layer is smaller in the central portion due to the acceleration of the flow stretching it. Axial bands of large and small particles have not formed yet. Instead, large particles dominate at the surface of the flowing layer, with a core of small particles having bulges where small particle bands will eventually appear.

The evolution of $\phi$ is more complex. In the upstream portion of the flow in Fig.~\ref{colorinit}(b), the upper thick region of large particles on the left ($0.02\lesssim x \lesssim 0.1)$ has a low density because the particles are accelerating and the packing is loose. As a consequence, the flow is stable, with lighter large particles above the heavier mixed region. However, the right side ($0.1\lesssim x \lesssim 0.2$) is unstable, with a heavier mixed region above lighter small particles. By the central region [Fig.~\ref{colorinit}(d)] the entire flow is unstable, with a dense upper layer above the small particle layer. Further downstream [Fig.~\ref{colorinit}(f)] the flow is decelerating, so the unstable condition persists  only on the right side, with the left side returning to a stable situation.

When axial segregation becomes stationary at long times, the $f_l$ fields at different points along the length of the flowing layer display distinct surface bands of large and small particles. The band of large particles becomes thick enough to fill the entire flowing layer in the upper and lower regions [Fig.~\ref{colorstatio}(a, e)] although it contains some small particles near the bottom of the flowing layer. It thins out in the central region [Fig.~\ref{colorstatio}(c)] with a core of small particles, consistent with previous experimental and simulation results~\cite{HillKakalios95, JainKhakhar01, Rapaport02, KhanMorris2004, Arndt05, TaberletNewey06, ChenLueptow11}. This also suggests that the overall fractions of large and small particles, $F_l$ and $F_s$, play a role in band formation in that there must be an adequate volume of small particles available for the small particle core to reach the surface, which we examine later in this paper.

The stationary volume fraction $\phi$ is similar. The upstream and downstream portions of the large particle band in Fig.~\ref{colorstatio}(b, f) have a thick layer of nearly pure large particles with low $\phi$ at the free surface and mixed particles below, making the band locally stable. In the central region [Fig.~\ref{colorstatio}(d)] the mixed layer with large $\phi$ is above a small particle layer with slightly lower $\phi$, but the small density inversion decreases the tendency for a Rayleigh-Taylor instability.  The small particle band that has emerged has negligible density inversion and, hence, no tendency for the RT instability.
Comparing the fields for $\phi$ in Figs.~\ref{colorinit} and~\ref{colorstatio}, it is evident that the evolution toward stationary bands results from the central region -- where the density gradient is initially unstable with upstream and downstream regions that are more stable -- to a situation after bands form where the density gradient is in a relatively stable condition for both the large and small particle bands along the entire length of the flowing layer.

\subsection{Recirculation rolls}

Now that we have established that the necessary condition for the granular Rayleigh-Taylor instability to occur is a layer of high-density material (mixed large and small particles) above a layer of low-density material (pure small particles), we can examine the flow that eventually results in band formation. This is best visualized as time-averaged particle displacement vectors in the $x$-$z'$ plane, where the flow is perpendicular to the plane of the image so that longitudinal rolls resulting from the instability are evident. To obtain the displacement vectors, the velocity throughout the entire domain at steady state is obtained by computing the velocity of particles on the nodes of a 3D subgrid, averaged over 100~s of physical time to ensure an adequately smooth velocity field. Particle displacements between two successive passes of particles through the $y'=0$~plane are computed by integrating the particle velocity based on the velocity field.
In other words, displacement vectors represent the change in particle position along a mean particle trajectory starting from the $y'=0$~plane, making a complete circuit through the flowing layer and solid body rotation zone, and then returning to the $y'=0$~plane.

\begin{figure}[hptb] 
\includegraphics[width=\linewidth]{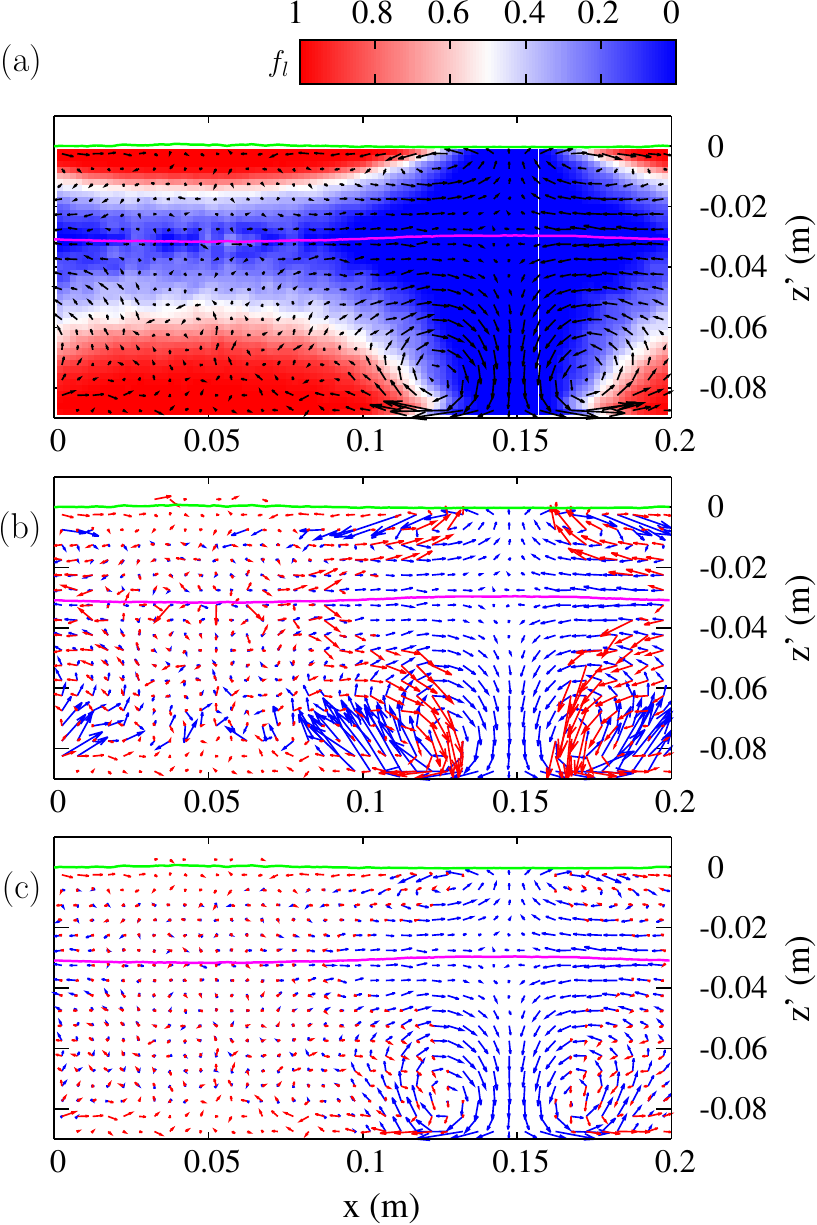}
\caption{Displacement vector maps in a plane perpendicular to the flowing layer and including the rotation axis (the $x$-$z'$ plane at $y'=0$) averaged over time $t=1000 \pm 50$~s, when the bands are steady for a periodic cylindrical tumbler with $L=100d$, $D=90d$, $\omega=15$~rpm, and filled to $V=0.5$ with equal volumes ($F_l=F_s=0.5$) of $d=2$~mm and $d_l=6$~mm spherical particles ($d_l/d=3$). Vector displacement lengths are magnified 3 times. (a) Particle displacement for combined large and small particles. (b) Separate particle displacements for the large (red) and small (blue) particles. (c) Particle displacement for each species multiplied by the local species fraction, $f_\mathrm{i}$. Only one particle species is present where only one color is shown. The green line indicates the free surface (based on $\phi=0.3$), and the magenta line delineates the flowing layer and static zone (based on $v'=0$).}
\label{dispmap}
\end{figure}

Figure~\ref{dispmap} displays three versions the average particle displacement that occurs upon one complete circuit through the system for steady-state bands at the same conditions as Fig.~\ref{colorstatio}(c, d) (250 tumbler rotations for $d_l/d=3$). Figure~\ref{dispmap}(a) shows the displacement vector map for all particles, large and small, superimposed on the $f_l$ field to show the large and small particle bands. The displacement vector lengths are magnified to make them more visible. First, it is evident that there is a pair of relatively strong longitudinally oriented recirculation rolls centered at $x=0.15$~m on either side of the small particle band. The pair of rolls is most evident at the bottom of the image in the solid body rotation portion of the granular bed. Of course, this is just the mirrored, stretched version of the recirculation rolls occurring in the flowing layer at the upper portion of the image at $x=0.15$~m.

The individual rolls of small and large particles are more evident in Fig.~\ref{dispmap}(b) where the displacements of large (red) and small (blue) particles are calculated as described above but separately for each species. Regions in which only one color of vectors appears have only one species present. Separating the vectors for the two species demonstrates that they follow different paths. This is most easily seen in the frozen, stretched recirculation rolls at the bottom of the image. 

To avoid overrepresentation of vectors in regions where few particles of a particular species are present, an alternative approach for the displacement vectors is shown in Fig.~\ref{dispmap}(c), where the displacement is multiplied by the local volume fraction for that species. The displacement vector map in Fig.~\ref{dispmap}(c) further demonstrates how the two species follow different paths. More importantly, it is evident that the small and large particle rolls overlap, and this overlap region contains both particle species. Furthermore, although counterintuitive, the overlapping small and large particle recirculation rolls have the same sense of rotation. This comes about due to size segregation in the mixed region where the recirculation rolls overlap. In this overlap region, the large particles rise and the small particles sink due to size segregation. As we explain in detail in our previous paper~\cite{DOrtonaLueptow24}, it is size segregation in this mixed region where the rolls overlap that drives the rolls themselves.

The result is a pair of small particle recirculation rolls within the small particle band. This pair of recirculation rolls brings small particles to the surface at $x=0.15$~m to reinforce the small particle band. At the same time, a pair of large particle recirculation rolls occurs at the edges of the large particle band. Each large particle recirculation rolls carries large particles away from the small particle band to enrich the large particle band. 

Recirculation rolls initiated by the granular RT instability occur in other situations, although they are somewhat different in nature from those occurring for size-bidisperse particles in rotating tumblers considered here. In the case of dense large particles and light small particles flowing down an incline~\cite{DOrtonaThomas20}, a pair of adjacent bands of small and  large particles (one wavelength) corresponds to two Rayleigh-B\'enard-like rolls extending through the entire depth of the flowing layer, with the upward motion associated with small particles and the downward motion associated with large dense particles. In contrast, in the present study, where the large and small particles have the same density, two adjacent bands correspond to four rolls that occur near the free surface. 

The recirculation rolls in a rotating tumbler are more similar to those for iso-density size-bidisperse particles flowing down an incline, with four rolls corresponding to one wavelength \cite{DOrtonaLueptow26}. In both cases, an upward flow occurs between small particle rolls at the middle of a small particle band. However, the large particle rolls are in opposite directions in the two cases. In the granular tumbler case considered here, segregation at the tilted mixed interface, where the large and small particle rolls overlap, drives the large particle roll upward at the interface, which is opposite to that for flow down an incline. The difference may lie in the relative strengths of the buoyancy-driven downward flow of the mixed particles and the segregation-driven upward movement of large particles, both occurring at the tilted mixed-particle interface. The buoyancy-driven  flow seems to dominate for the incline, whereas segregation dominates in the rotating tumbler. However, the differences are subtle and deserve further investigation. {\blue For instance, additional insight may be possible by varying the particle size ratio, relative densities, or friction coefficient, which are likely to impact the sense of rotation and the overall geometry of the recirculation cells.}

\FloatBarrier

\section{Effect of tumbler and particle parameters}
Axial segregation depends on several parameters, including tumbler geometry and the sizes of the particles. In this section, we explore how these parameters, and others, affect axial segregation as characterized using the segregation index, $I_\mathrm{seg}$, and the associated particle volume fraction profiles that are related to the appearance of the granular Rayleigh-Taylor instability.

\subsection{Tumbler length}

Since the simulations can be computationally time-consuming for long tumblers, we take advantage of the periodicity of the axial segregation to simulate a single pair of large and small particle bands using periodic boundary conditions, as in the previous section.  Furthermore, previous results indicate that optimal axial segregation occurs for size ratios in the range $2\le d_l/d \le 3.5$ in terms of both the time for bands to fully form and the degree of segregation~\cite{DOrtonaLueptow24} (see also Figs.~\ref{isvsdl} and \ref{profile}).  Hence, we consider detailed aspects of band formation for $d_l/d=2$ and 3. To further accelerate the formation of bands, we use a sinusoidal variation in the initial volume fraction of the particle species as described in previous sections. 

\begin{figure}[hptb] 
\includegraphics[width=\linewidth]{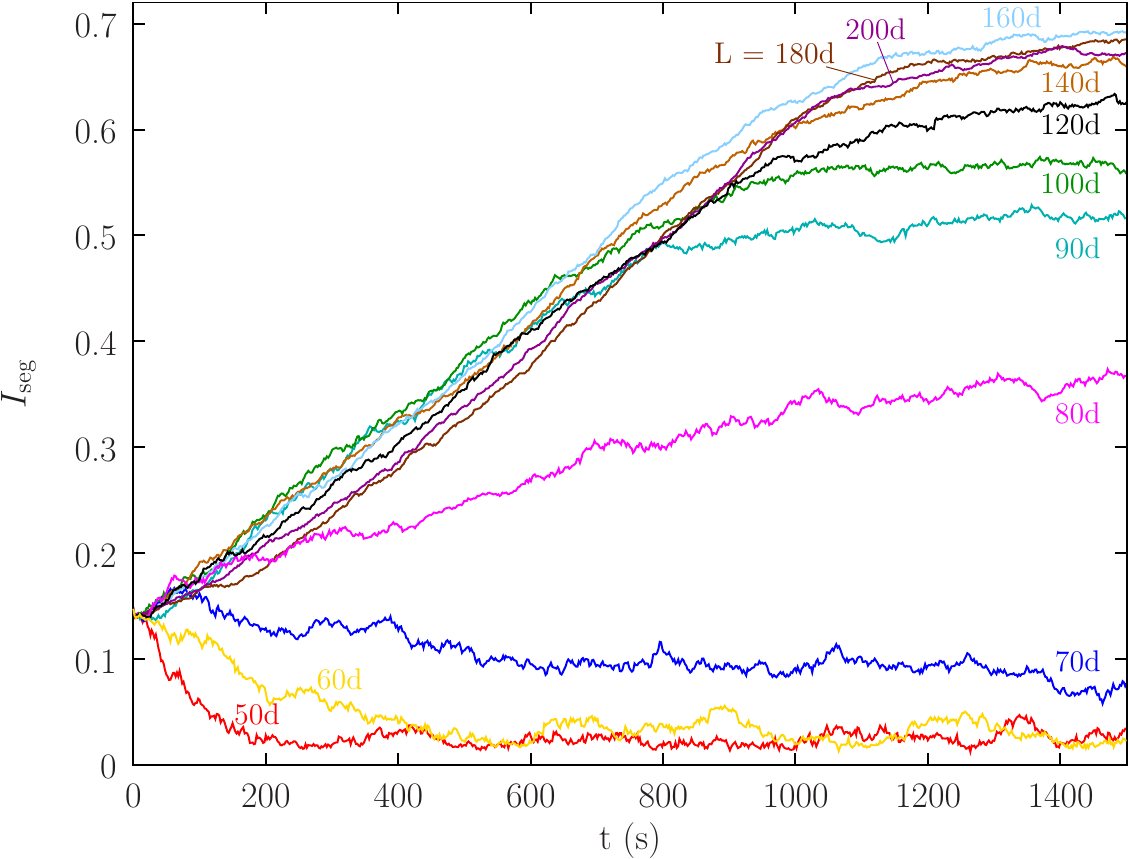}
\caption{Effect of periodic tumbler length, $50d \leq L \leq 200d$, on the time evolution of the axial index of segregation, $I_\mathrm{seg}$, starting with a sinusoidal variation in the initial species volume fraction ($\Delta f_l=0.3$) in a tumbler with $D=90d$, filled to $V=0.5$
with equal volumes ($F_l=F_s=0.5$) of 2~mm and 4~mm spherical particles ($d_l/d=2$), and rotating at $\omega=15$~rpm. 
}
\label{isvsW}
\end{figure}

Consider first the effect of varying the axial length of the periodic tumbler from $L=50d$ to $200d$ on axial segregation for $d_l/d=2$, shown in Fig.~\ref{isvsW}. The axial segregation index, $I_\mathrm{seg}$, starts with a small value at $t=0$, reflecting the initial sinusoidal variation in the relative species fractions. $I_\mathrm{seg}$ increases with time, indicating increasing axial segregation, provided that $L \ge 90d$, noting that here the tumbler diameter is $D=90d$. That band formation occurs when the band wavelength, which here is controlled by the periodic length of the tumbler, is about the same as the tumbler diameter is consistent with experimental results for long tumblers showing that the wavelength of the bands is about $1D$~\cite{CharlesMorris06, FiedorOttino06, JuarezLueptow08, JuarezOttino10} before bands eventually merge at very long times. The plateau for $I_\mathrm{seg}$ becomes increasingly higher for $L\ge 80d$, indicating stronger axial segregation, up to about $L=140d$, after which increasing $L$ has little impact on the plateau for $I_\mathrm{seg}$. 
The time evolution of $I_{\rm seg}$ is similar for $L \ge 90d$, showing no dependence on $L$. In contrast, axial segregation is slower for $L=80d$, and for $L\leq 70d$, $I_\mathrm{seg}$ decreases, indicating remixing of the initial variation in the species volume fraction with no tendency toward band formation at all.

Before going on, we briefly address the validity of using a sinusoidal perturbation to promote faster development of the bands than can occur starting with a fully mixed initial condition in a longer tumbler. To do so, we compare the time evolution of $I_{\rm seg}$ for $d_l/d=2$ in an $L=100d$ short tumbler having periodic boundary conditions and an initial concentration variation of $\Delta f_l=0.3$ (see Fig.~\ref{isvsW}) with that for a 1~m long periodic tumbler, corresponding to $L=500d$, starting with a fully mixed initial condition of $\Delta f_l=0$ for the same particle size ratio $d_l/d=2$ (see Fig.~\ref{isvsdl} and \cite{DOrtonaLueptow24}). First, the steady-state value of $I_{\rm seg}$ in Fig.~\ref{isvsW} at $L=100d$ in the short tumbler  matches the steady state value for $d_l/d=2$ in Fig.~\ref{isvsdl} in the long tumbler.  However, the shorter tumbler reaches the same value in about half the physical time, demonstrating that using an initial variation in the species volume fraction in a short tumbler does not affect the stationary value for $I_\mathrm{seg}$. A drawback of using a short tumbler with $\Delta f_l\ne0$ is that the initial exponential growth of $I_\mathrm{seg}$ evident in Fig.~\ref{isvsdl}(b) does not appear. This is because $I_\mathrm{seg}=0.14$ at $t=0$ for the short tumbler cases, whereas for the long tumbler in Fig.~\ref{isvsdl}, this value for $I_\mathrm{seg}$ is not reached until much later. 
For much of the remainder of this paper, a short periodic tumbler with $L=100d = 1.11 D = 20$~cm and an initial perturbation is used, since the time evolution of $I_{\rm seg}$ and the plateau match those for the long tumbler without perturbation. 

It is interesting to note that in short tumblers (Fig.~\ref{isvsW}), the highest segregation index is reached for a tumbler length, and thus a wavelength $\lambda$, of $L\simeq 160d$, longer than the tumbler diameter. In a long tumbler of $L=500d$, this should result in approximately 3 bands, whereas 5 bands are obtained (see Fig.~\ref{d3w500}), leading to a wavelength of $\lambda = 100d$.
The higher values of $I_\mathrm{seg}$ for larger $L$ are consistent with the tendency toward coarsening after long times that has been observed experimentally~\cite{Frette97, HillCaprihan97, ChooMorris97, ChooMorris98, FiedorOttino03, FiedorOttino06, JuarezLueptow08, JuarezOttino10}.

\subsection{Particle size ratio}

\begin{figure}[hptb] 
\includegraphics[width=\linewidth]{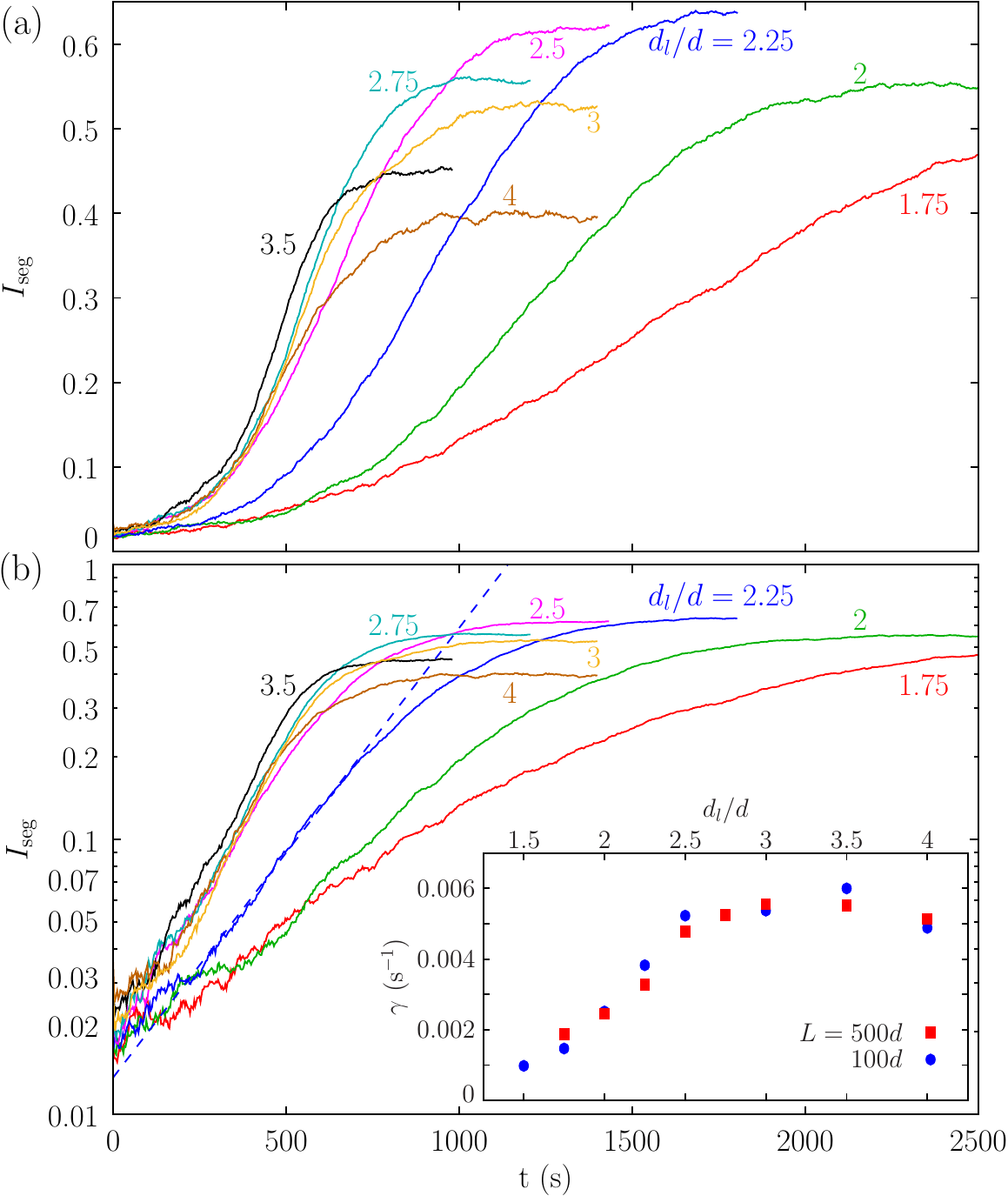}
\caption{(a) Time evolution of $I_\mathrm{seg}$ for
$d_l/d=1.75$ to 4 with $L=500d$, $D=90d$, $\omega=15$~rpm, and filled to $V=0.5$
with $d=2$~mm small particles and large particles of diameter $d_l$ ($F_l=F_s=0.5$). (b) Same curves on a semi-log plot to show the initial exponential growth. The dashed line indicates how growth rate $\gamma$ is measured. Inset shows $\gamma$ vs. $d_l/d$ for two tumbler lengths. See Fig.~S1 in SM for the time evolution of  $I_\mathrm{seg}$ for $L=100d$. 
}
\label{isvsdl}
\end{figure}

The particle size ratio, $d_l/d$, affects the evolution of the axial segregation index, $I_\mathrm{seg}$, in a 1 m long tumbler, as shown in Fig.~\ref{isvsdl}. Here the size ratio, $d_l/d$, is varied by changing $d_l$ while keeping $d=2$~mm. The time evolution of axial segregation for size ratios $d_l/d\geq2.5$ is much faster than for smaller size ratios, taking only about 500--1000 s (125 to 250 tumbler rotations). 
This is not surprising, since a large particle size ratio gives a high packing fraction. Although the maximum segregation index of $I_\mathrm{seg}\gtrsim 0.6$ occurs for both $d_l/d=2.25$ and $d_l/d=2.5$, the $d_l/d=2.5$ case reaches the plateau more quickly than the $d_l/d=2.25$ case. 

Figure~\ref{isvsdl}(b) shows the same curves on a semi-log plot to demonstrate the initial exponential growth, $I_{\rm seg}\sim {\rm e}^{\gamma t}$. The dashed line indicates the typical portion of the plot used to measure the growth rate $\gamma$. The inset in Fig.~\ref{isvsdl}(b) shows how the growth rate increases with $d_l/d$ but eventually reaches a plateau {\blue and even seems to slow for $d_l/d=4$. Although larger size ratios could not be considered due to computational limitations, 
it is possible that this is a result of the mechanism for segregation changing as the size ratio increases. Indeed, for $d_l/d>3$, granular size segregation becomes less efficient \cite{Thomas00,DOrtonaThomas18} and even vanishes for $d_l/d\gtrsim 6.46$, where small particles are so small that they can segregate through spaces between even static large particles~\cite{Dodds80, Vyas24}.} Also included in the inset are growth rates from simulations of short tumblers ($L=100d$) with $\Delta f_l=0.1$ (to capture the exponential growth portion of the curve), which match the growth rate results for long tumblers ($L=500d$), further confirming the validity of the short tumbler approach.

\begin{figure}[hptb] 
\includegraphics[width=0.9\linewidth]{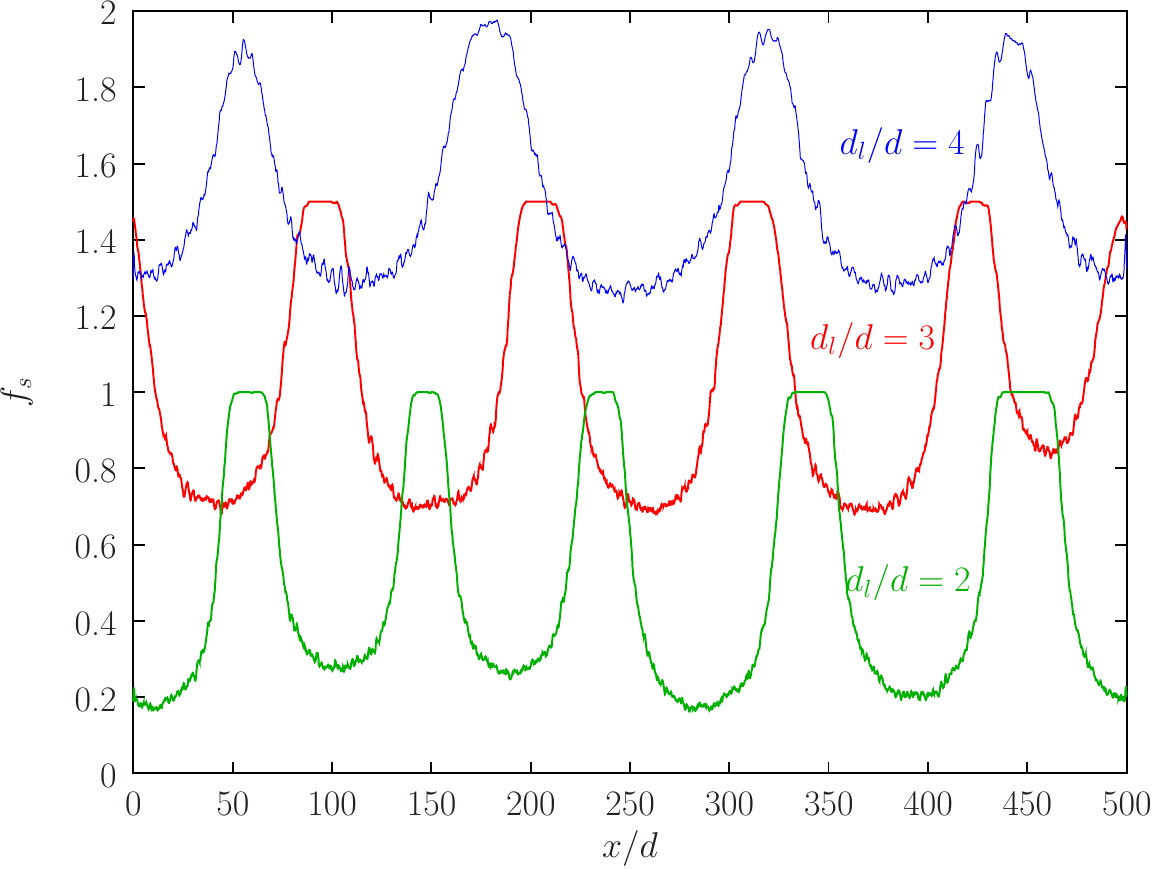}
\caption{Small particle volume fraction, $f_s$, along the length of the tumbler at steady-state 
with $L=500d$, $D=90d$, $\omega=15$~rpm, and $V=0.5$
for $d=2$~mm small particles, with $d_l/d=2$ to 4 ($F_l=F_s=0.5$). The $d_l/d=2$ curve is 
correctly placed with respect to the $f_s$ scale; curves for other size ratios are shifted vertically by 0.5 and 1.0 for clarity. See Fig S2 in SM for equivalent profiles at the time when $I_\mathrm{seg}=0.1$.
}
\label{profile}
\end{figure}

Figure~\ref{profile} shows the volume fraction of small particles, $f_\mathrm{s}$, along the length of the $L=500d$ tumbler when axial segregation is stationary. Four or five bands of small and large particles are evident for all size ratios, in agreement with experimental results \cite{JuarezLueptow08}. 
Thus, for the $L=500d$ tumbler, the wavelength for a pair of bands is $\lambda \sim100d$, consistent with our approach using a short tumbler of $L=100d$ to reduce simulation time. For size ratios $d_l/d=2$ and 3, the small particle bands are nearly pure ($f_\mathrm{s}=1$, noting the vertical offset in the figure), while bands between the pure small particle bands consist of a core of small particles surrounded by mostly large particles, resulting in $f_\mathrm{s}\approx 0.2$. At $d_l/d=4$, the small particle band has $f_\mathrm{s}<1$, indicating that some large particles are mixed into the small particle bands.

\begin{figure}[hptb]
\includegraphics[width=\linewidth]{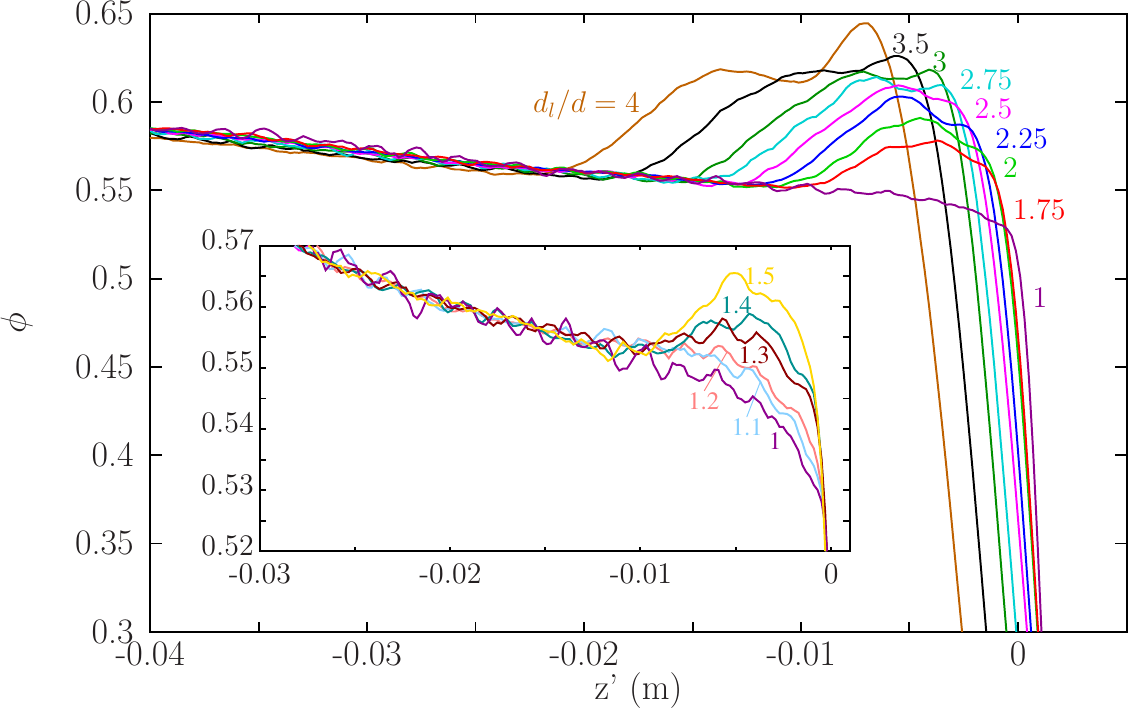}
\caption{Depthwise profiles of $\phi$ for $1 \leq d_l/d \leq 4$ in a 1~cm thick slice centered at $y'=0$ with $D=90d$, $\omega=15$~rpm, $d=2$~mm, $V=0.5$ ($F_l=F_s=0.5$), and $L=50d$ to suppress the RT instability. Upper portions of the profiles are shown ($z'\le -0.04$~m).       See Fig.~S3 in SM for entire profiles. Inset shows $\phi$ for $1\le d_l/d \le 1.5$. 
}
\label{volfracb}
\end{figure}

For an RT instability, either fluid or granular, the growth rate depends on the density ratio between the layers, which for size-bidisperse particles having the same density can only occur due to a denser mixed layer packing. To confirm this, 
Fig.~\ref{volfracb} shows $\phi$ as a function of depth in the granular flow ($z'$) for $1\leq d_l/d \leq 4$ in a short tumbler with $L=50d$ to prevent any axial segregation. Indeed, the bump in the value of $\phi$ near the flowing layer surface increases with $d_l/d$, consistent with the growth rate in the inset of Fig.~\ref{isvsdl}(b) as $d_l/d$ increases from 1.75 to 3.5. 
The growth rate seems to approach $\gamma=0$ for smaller $d_l/d$, suggesting that axial segregation should not occur, as would be expected as the difference in particle sizes becomes negligible. However, Fig.~\ref{volfracb} shows that even for very small particle size ratios, a bump in $\phi$ occurs where the particles are mixed. 

\begin{figure}[hptb] 
\includegraphics[width=\linewidth]{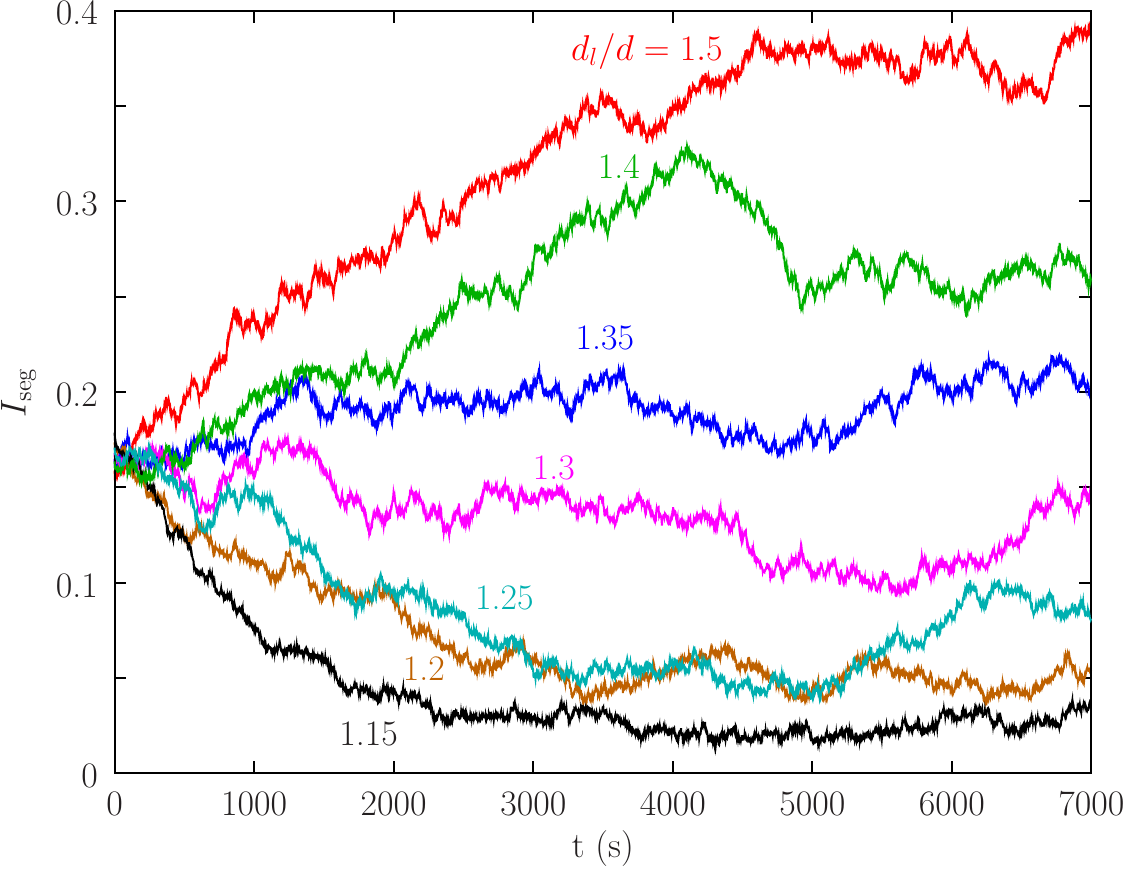}
\caption{Effect of particle size ratio $d_l/d$ on the time evolution of $I_{\rm seg}$ with $L=100d$, $D=90d$, $\Delta f_l=0.3$, $\omega=15$~rpm, and filled to $V=0.5$
with $d=2$~mm small particles and large particles of diameter $d_l$ ($F_l=F_s=0.5$). 
}
\label{seuildl}
\end{figure}

To consider small size ratios further, we use a short tumbler with $\Delta f_l=0.3$. Figure~\ref{seuildl} shows the time evolution of $I_\mathrm{seg}$ for $1.15 \leq d_l/d \leq 1.5$. 
As $d_l/d$ decreases, the tendency for segregation decreases. In fact, for $d_l/d \le 1.3$, $I_\mathrm{seg}$ decreases with time as the initial species volume fraction variation is damped out. 
Nevertheless, $I_\mathrm{seg}$ approaches a finite nonzero value rather than going to zero, suggesting a slight tendency toward axial segregation even for $d_l/d = 1.15$. In fact, plotting the value of $I_\mathrm{seg}$ at the plateau versus $d_l/d$ results in a linear relationship, with $I_\mathrm{seg}$ extrapolating to zero at $d_l/d \approx 1.1$ (see Fig.~S4 in the SM), which happens to be the size ratio where the distributions of particle sizes used to avoid a close-packed ordered structure begin to overlap and for which segregation should be negligible. 
Hence, there is no specific size ratio threshold required to induce axial segregation. However, when the magnitude of axial segregation as quantified by $I_{\rm seg}$ is low, {\blue $I_{\rm seg} \lesssim 0.2$, axial segregation consists of} subsurface bulges of the core of small particles that are not visible at the free surface. {\blue Scattered small particles at the free surface that might be interpreted as the first sign of a weak surface band become visually evident as $d_l/d$ increases above 1.35, which corresponds to $I_{\rm seg} \simeq 0.2$ in Fig.~\ref{seuildl} (see Fig. S5 in the SM).}

\subsection{Density ratio} 

 \begin{figure}[hptb] 
\includegraphics[width=0.9\linewidth]{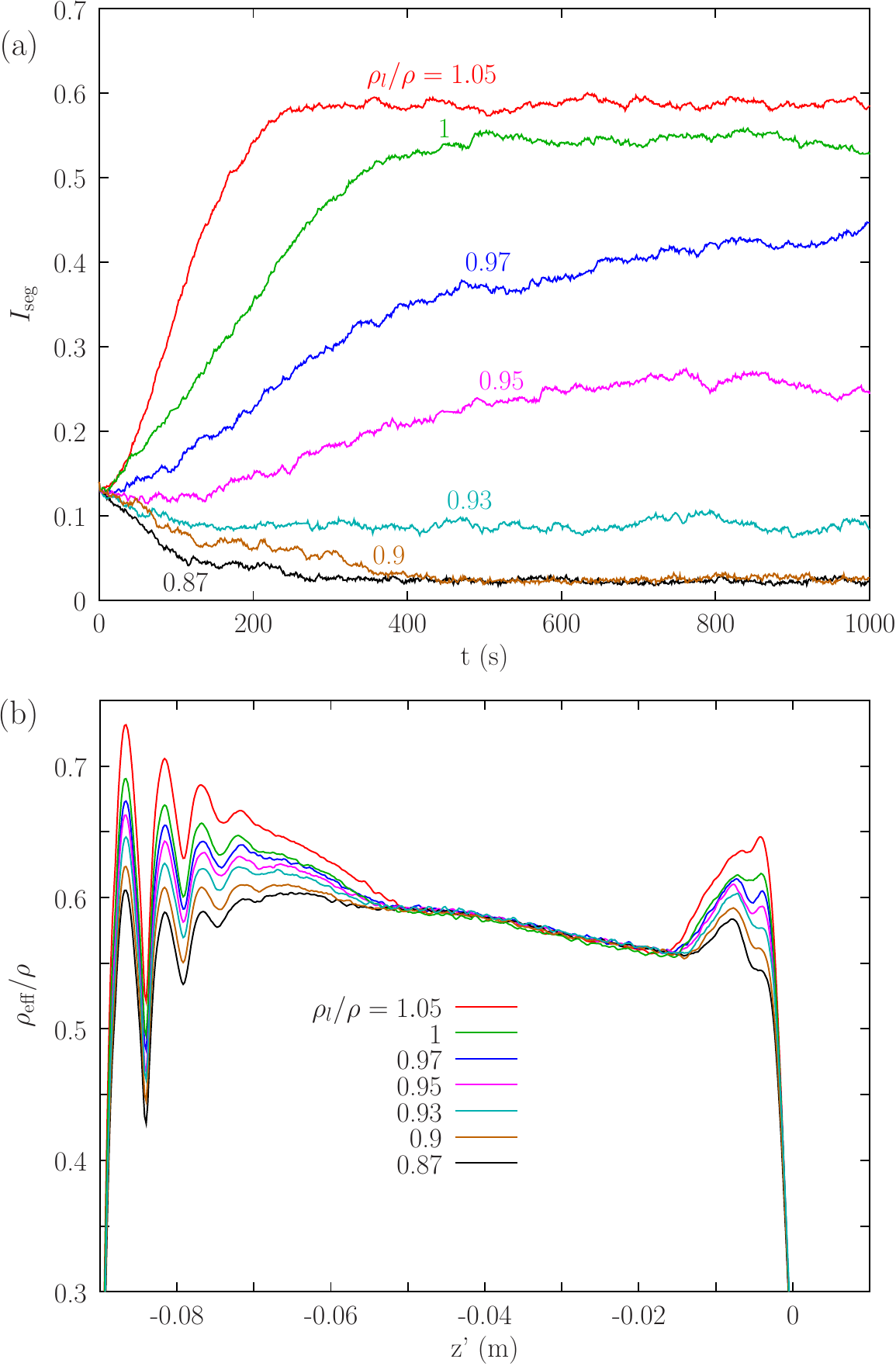}
\caption{(a) Time evolution of $I_{\rm seg}$ for 
density ratios $0.87 \leq \rho_l/\rho \leq 1.05$ with $L=100d$, $D=90d$, $\Delta f_l=0.3$, $\omega=15$~rpm, $d_l/d=3$ ($d=2$~mm), and filled to $V=0.5$ with $F_l=F_s=0.5$. (b) Effective density profile in a 1~cm thick slice centered at $y'=0$ with $L=50d$ to suppress the RT instability.  
}
\label{seuildens}
\end{figure}

{\blue To further confirm that band formation is initiated by the RT instability, we} vary the ratio of the density of the large particles to that of the small particles, $\rho_l/\rho$, keeping the size ratio at $d_l/d=3$. The idea here is that small decreases in $\rho_l/\rho$ at the same $\phi$ makes the mixed layer less dense, {\blue which should decrease} the propensity for the granular RT instability and slow {\blue down or eliminate}  band formation, while small increases in $\rho_l/\rho$ makes the mixed layer denser, {\blue which should lead} to faster band formation. 

Figure~\ref{seuildens}(a) shows the evolution of $I_\mathrm{seg}$ for different values of $\rho_l/\rho$ in a short periodic tumbler with  $\Delta f_l=0.3$. For decreasing $\rho_l/\rho$, axial segregation occurs more slowly and reaches a lower plateau, just as would be expected for a smaller effective density of the mixed layer, {\blue corresponding to reduced likelihood of the RT instability}. In fact, for $\rho_l/\rho\leq 0.9$, the axial segregation decreases and particles remix, leaving no trace of the initial sinusoidal variation in species fractions. For $\rho_l/\rho=0.93$, $I_{\rm seg}$ decreases but does not vanish. The species fractions retain a sinusoidal variation with an amplitude that is slightly smaller than the initial condition. For $\rho_l/\rho=1.05$, the axial segregation is faster than for $\rho_l/\rho=1$ and reaches a higher plateau, {\blue again} just as expected {\blue for a RT instability-initiated bands}.

Figure~\ref{seuildens}(b) shows the profile of the relative density, $\rho_{\rm eff}/\rho$, where $\rho_{\rm eff}=\rho_l f_l + \rho_s f_s$. The mixed-layer density bump is reduced for $\rho_l/\rho<1$, leading to reduced propensity for axial segregation, while for $\rho_l/\rho>1$, the increased mixed layer density leads to more effective axial segregation. 
Figure~\ref{seuildens}(b) highlights the underlying mechanism: since the excess density bump responsible for the RT instability primarily originates from the increased volume fraction, $\phi$, associated with mixed particles, a slightly lower large particle density is not sufficient to counteract it. However, it appears that a minimal density gradient is required to trigger the RT instability, since no instability occurs for a density ratio $\rho_l/\rho \leq 0.9$, even though a small excess density bump is observed in Fig.~\ref{seuildens}(b). {\blue Maps of $\rho_{\rm eff}/\rho$ in the $x-z'$~plane confirm that the amplitude of the waviness related to the RT instability at the interface between large and small particles decreases as $\rho_l/\rho$ decreases to the point that it is negligible for $\rho_l/\rho = 0.87$.}
\FloatBarrier

\subsection{Particle size and tumbler diameter}

The ratio of the tumbler diameter to the particle size has been shown to be a critical parameter for the onset of axial segregation.  Previous results based on an extensive set of experiments~\cite{AlexanderMuzzio04} indicate that axial segregation evident as visible bands at the surface of the flowing layer never occurs for $D \leq [20(d_l/d+1)d]$ and  always occurs for $D \geq [27.5(d_l/d+1)d]$, noting that these relations are based on experiments with $2 \lesssim d_l/d \lesssim 6.5$, but clearly cannot apply as $d_l/d \rightarrow1$, for which no radial or axial segregation occurs. 
The simulation results in Fig.~\ref{isvsW} for $d_l/d=2$ are consistent with this experimental criterion, which predicts that axial segregation should first be evident for $60d \leq D \leq 82.5d$, noting that $I_{\rm seg}$ detects subsurface bulges of the small particle core.

\begin{figure}[hptb]
\includegraphics[width=\linewidth]{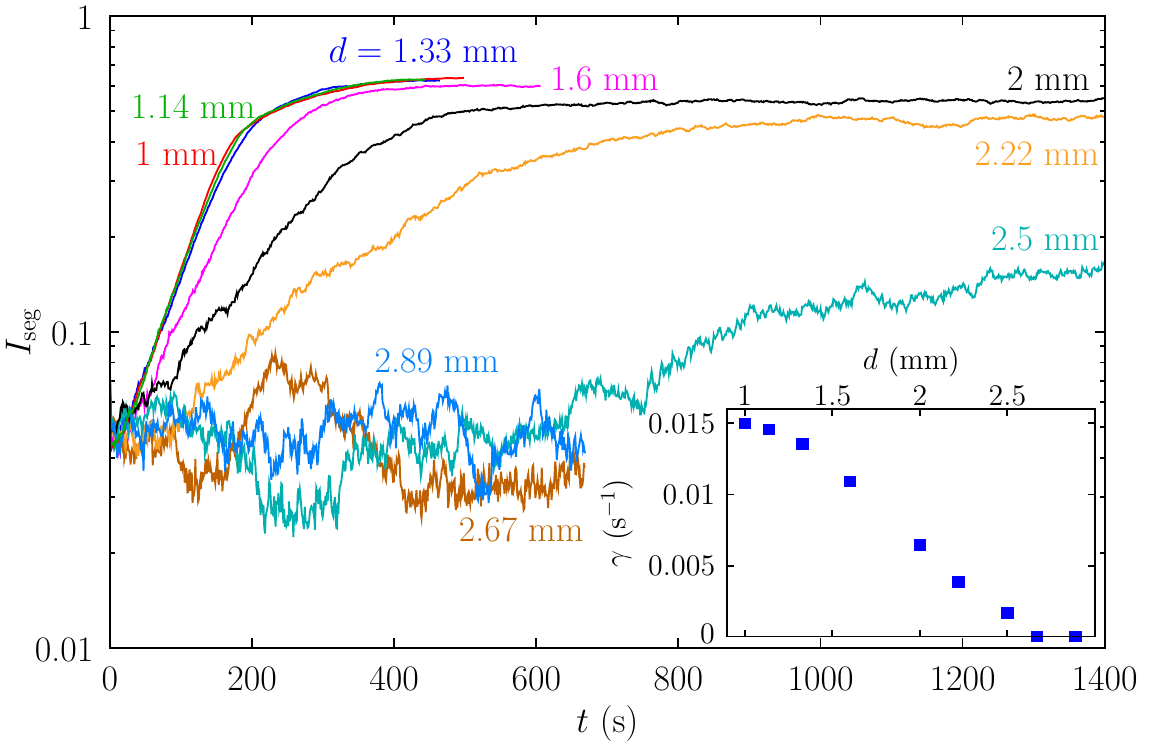}
\caption{Time evolution of $I_{\rm seg}$ for $1 \leq d \leq 2.89$~mm (corresponding to $180d \geq D \geq 62d$) on a semi-log scale, with $L=100d$, $d_l/d=3$, $\omega=15$~rpm, and $\Delta f_l=0.1$. See Fig.~S6 in SM for the plot on a linear scale. Inset: Growth rate $\gamma$ versus small particle size $d$. 
}
\label{Isegvsd}
\end{figure}

 To test this further, we consider a particle size ratio of $d_l/d=3$ and tumbler diameter $D=0.18$~m while varying the particle size. For these conditions, the minimum criterion for visible axial segregation bands~\cite{AlexanderMuzzio04} corresponds to $D \geq 80d$, or $d \leq 2.25$~mm. Figure~\ref{Isegvsd} shows the time evolution of $I_{\rm seg}$ for $d=1$~mm to 2.89~mm. For $d \leq 2.22$~mm, $I_{\rm seg}$ increases and bands form. However, axial segregation is weak or does not occur at all for $d \geq 2.5$~mm.  
Growth rates, $\gamma$, extracted from the time evolution of $I_{\rm seg}$ and shown in the inset of Fig.~\ref{Isegvsd}, decrease with increasing particle size and vanish around $d=2.5$~mm. The results in Figs.~\ref{isvsW} and~\ref{Isegvsd} not only demonstrate the dependence of axial segregation on $D/d$ but also confirm the validity of our simulations when compared to experimental results~\cite{AlexanderMuzzio04}.

While the dependence of axial segregation on particle size has been known for some time~\cite{AlexanderMuzzio04}, it is interesting to reanalyze this dependence through the lens of the RT instability. Earlier in this paper, an increase in the growth rate was linked to an increase in the density of the mixed zone. However, the profiles of $\phi$ (see Fig.~S7 in the SM) for $d_l/d=3$ with particle sizes ranging from $d=1$~mm to 2.5 mm -- spanning from an unstable case with axial segregation to a stable one having no segregation -- are very similar, with comparable density inversions. This result is expected because the increase in $\phi$ for mixed particles is a purely geometric phenomenon that depends solely on $d_l/d$, which remains constant here. Consequently, the increase in $\gamma$ for smaller particles is not linked to an increase in $\phi$ but is most likely linked to the flowability of the particles. This result is analogous to the RT instability in viscous fluids, where the growth rate is inversely proportional to viscosity \cite{Chandrasekhar61,Selig65,Ramberg68,DOrtonaMartinand25}. However, two differences from fluids should be noted. For larger particles, the growth rate not only decreases but vanishes for $d \gtrsim 2.5$~mm, indicating the existence of a threshold. Furthermore, the growth rate saturates as particle size decreases. These results suggest that other mechanisms may come into play, which deserves further study.

\begin{figure}[hptb] 
\includegraphics[width=\linewidth]{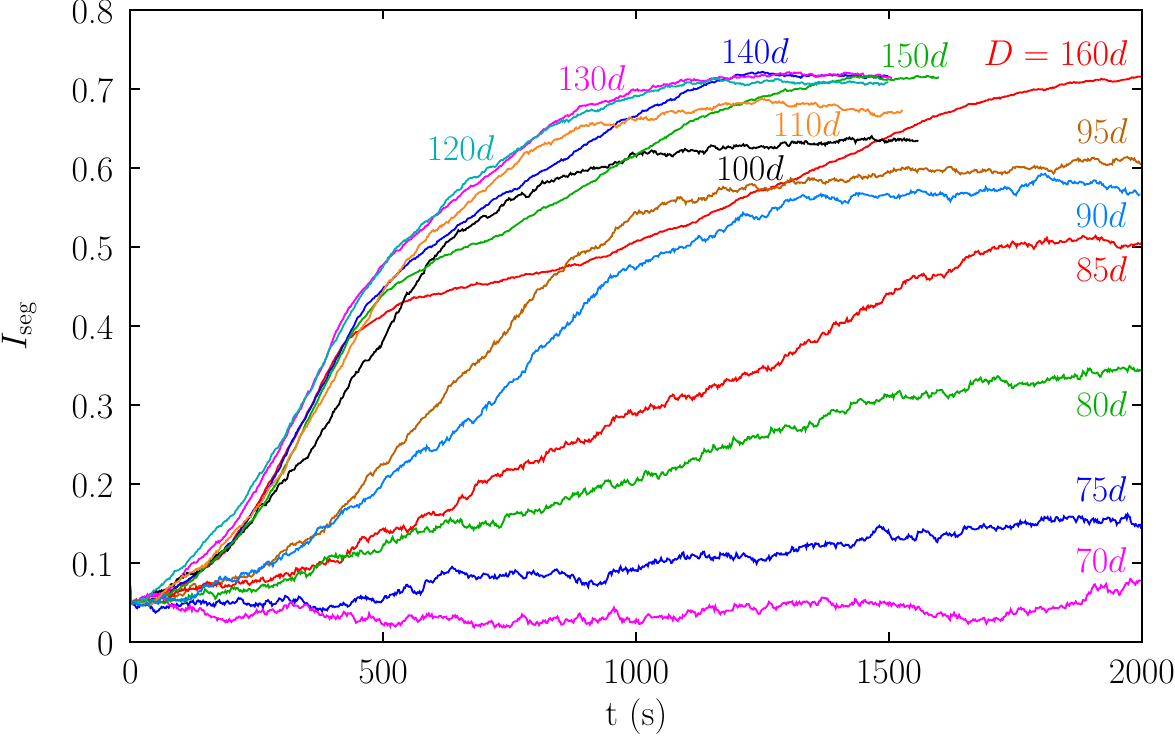}
\caption{Effect of periodic tumbler diameter, $70d \leq D \leq 160d$, on the time evolution of $I_\mathrm{seg}$ with $L$ adjusted to keep $L/D\simeq 1.1$, $\Delta f_l=0.1$, $\omega=15$~rpm, $d_l/d=2$ ($d=2$~mm), and filled to $V=0.5$ with $F_l=F_s=0.5$. Semi-log plots are included as Fig.~S8 in the SM. See Fig.~S10 in the SM for the same graph with $d_l/d=3$.
}
\label{d2wxxxtime}
\end{figure}

Rather than keeping the tumbler diameter $D$ constant and changing the particle size $d$ as in Fig.~\ref{Isegvsd}, here we do the opposite. Figure~\ref{d2wxxxtime} shows the time evolution of $I_\mathrm{seg}$ for a short periodic tumbler with  $L/D\simeq 1.1$ ($1.10 \leq L/D \leq 1.13$) and $d_l/d=2$. The axial segregation grows stronger in terms of both the time evolution of band formation and the degree of segregation as $D/d$ increases, up to $D=130d$. For larger $D/d$, the growth of $I_{\rm seg}$ slows down around $I_{\rm seg}\simeq 0.4$ but eventually plateaus at a similarly high level.
Consistent with the criterion from experiments~\cite{AlexanderMuzzio04}, axial segregation first becomes evident in the range $60d \leq D \leq 82.5d$ based on $d_l/d=2$.
Results for the long periodic tumbler case 
are similar (see Fig.~S9 in the SM~\cite{SM}). 

\begin{figure}[hptb] 
\includegraphics[width=\linewidth]{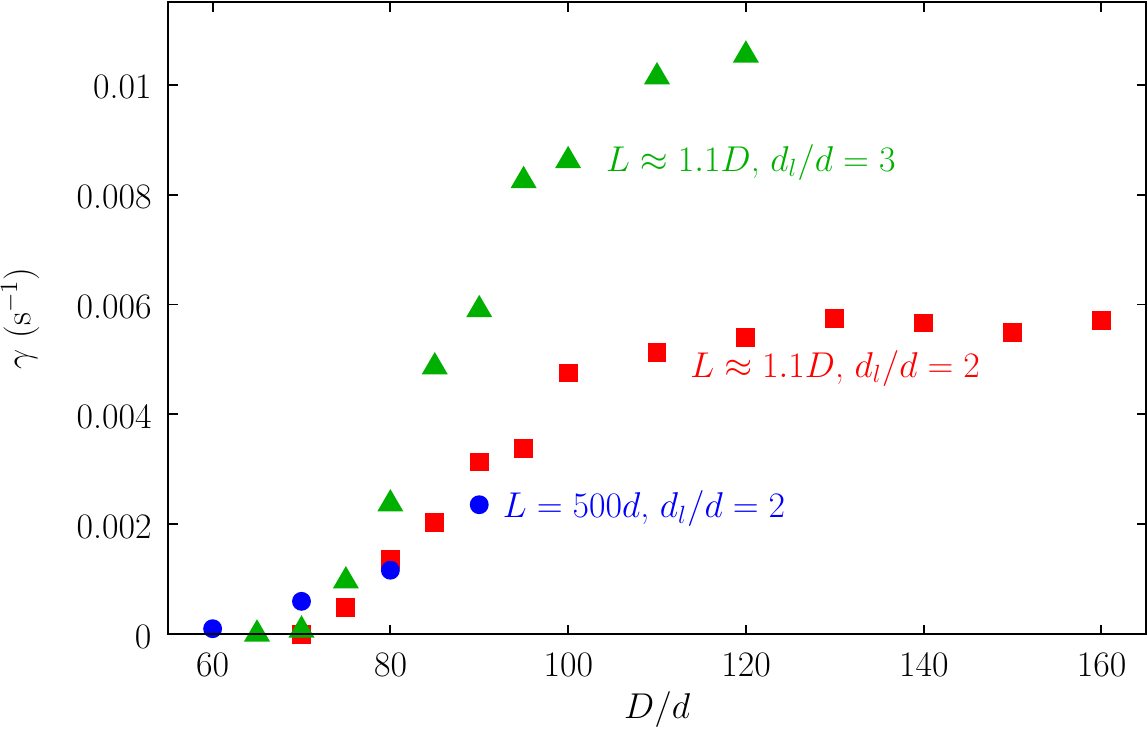}
\caption{Growth rate $\gamma$ versus tumbler diameter with other parameters as in Fig.~\ref{d2wxxxtime} for {\blue $d_l/d=2$ ($L \approx 1.1 D$) for the red squares, and $L=500d$, corresponding to $5.55 \leq L/D \leq 8.33$, for the blue circles}, and $d_l/d=3$ {\blue ($L \approx 1.1 D$)} for the green triangles.} 
\label{growthratevsD}
\end{figure}

Figure~\ref{growthratevsD} presents the growth rates measured from the initial part of the $I_{\rm seg}$ curves as a function of tumbler diameter $D$ for short {\blue ($L \approx 1.1 D$)} and long tumblers {\blue ($L = 500 d$, so that $5.55 \leq L/D \leq 8.33$)} with particle size ratio $d_l/d=2$, and for short tumblers with $d_l/d=3$. The growth rate increases monotonically and saturates for the largest tumbler diameters. This confirms that the growth rate is not affected by the initial sinusoidal perturbation in short tumblers and shows that the growth rates for $d_l/d=3$ are about twice those for $d_l/d=2$, independent of the tumbler diameter.

The dependence of axial segregation on $D/d$ in Fig.~\ref{growthratevsD} can be partially explained in terms of the RT instability. For a fluid RT instability, the growth rate is proportional to the density difference and to the wavelength of the instability, which itself is proportional to the layer thickness~\cite{DOrtonaMartinand25, Ramberg68, Selig65}. We observe that the density difference is nearly unaffected by the tumbler diameter (see Fig.~S11 in the SM). One therefore expects the growth rate of the granular instability to increase with the flow thickness, which in turn increases with the tumbler diameter, as is the case in Fig.~\ref{growthratevsD}. However, there are differences between the RT instability leading to axial band formation considered here and both the fluid RT instability and the granular RT instability on an inclined plane with size-bidisperse particles of equal density~\cite{DOrtonaLueptow26}. First, a threshold exists, as neither the instability nor axial segregation develops below $D/d=70$, evident as $\gamma=0$ in Fig.~\ref{growthratevsD}. Second, $\gamma$ saturates for large tumbler diameter. These differences are most likely related to the specific geometry of the flow in a rotating tumbler, {\blue perhaps related to the distance along the length of the flowing layer for segregation to be complete (see the variation in $\phi$ with position along the length of the flowing layer in Figs.~\ref{colorinit} and \ref{colorstatio})}.

\subsection{Particle species fraction}

\begin{figure}[hptb] 
\includegraphics[width=\linewidth]{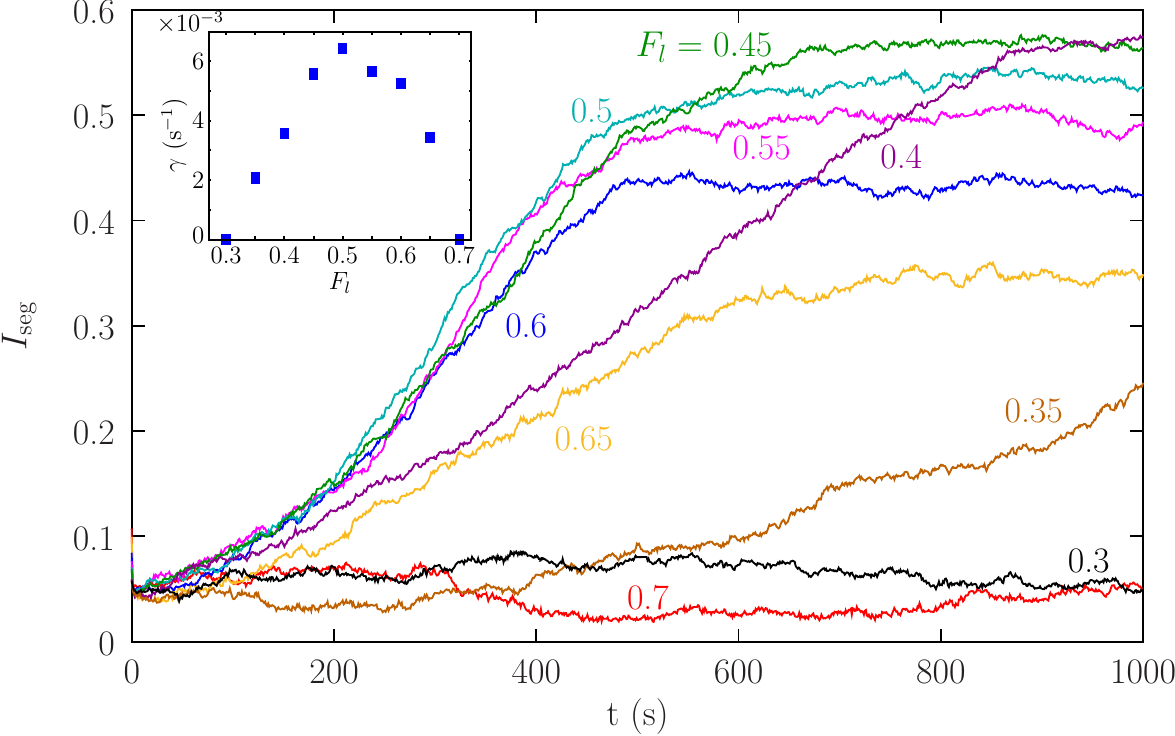}
	\caption{Time evolution of $I_{\rm seg}$ for large particle fraction $0.3 \leq F_l \leq 0.7$ with $L=100d$, $D=90d$, $\Delta f_l=0.1$, $\omega=15$~rpm, $d_l/d=3$ ($d=2$~mm), and filled to $V=0.5$. Inset: growth rate $\gamma$ versus $F_l$. 
    } 
\label{isegvsfs}
\end{figure}

In previous sections and in the majority of published studies on axial segregation, the volume fractions of large and small particles are equal, $F_l=F_s=0.5$. Yet, Fig.~\ref{isegvsfs} shows how these fractions strongly affect axial segregation, measured in terms of the time evolution of $I_{\rm seg}$. $F_l=0.5$ is the most unstable in terms of reaching its plateau most quickly, but $F_l=0.45$ and 0.4 lead to slightly more segregation, although the growth is much slower for $F_l=0.4$ (see the inset of Fig.~\ref{isegvsfs}). In contrast, no axial segregation occurs for species fractions that are farthest from equal volume fractions, $F_l=0.7$ and $F_l=0.3$. The maximum growth rate occurring for $F_l \approx 0.5$ and the symmetry of the value for $\gamma$ about this value, both of which are quite evident in the inset of Fig.~\ref{isegvsfs}, suggest that optimal evolution of axial segregation may be related to having adequate volumes of both particle sizes available. We simply note that for $F_l=0.5$, the thickness of the dense mixed layer occupies about half of the flowing layer (see Fig.~S12 in the SM), leading to a maximum growth rate for the RTI \cite{DOrtonaMartinand25,Ramberg68,Selig65}. 
This result may appear to contradict the case of the inclined plane, where maximum segregation is observed for $F_l\simeq 0.2$ \cite{DOrtonaLueptow26}. The difference simply arises from the fact that, in a cylinder, as large particles locate at the periphery, the thickness of the large particle layer is reduced compared to the small one.

\subsection{Fill level}

\begin{figure}[hptb] 
\includegraphics[width=\linewidth]{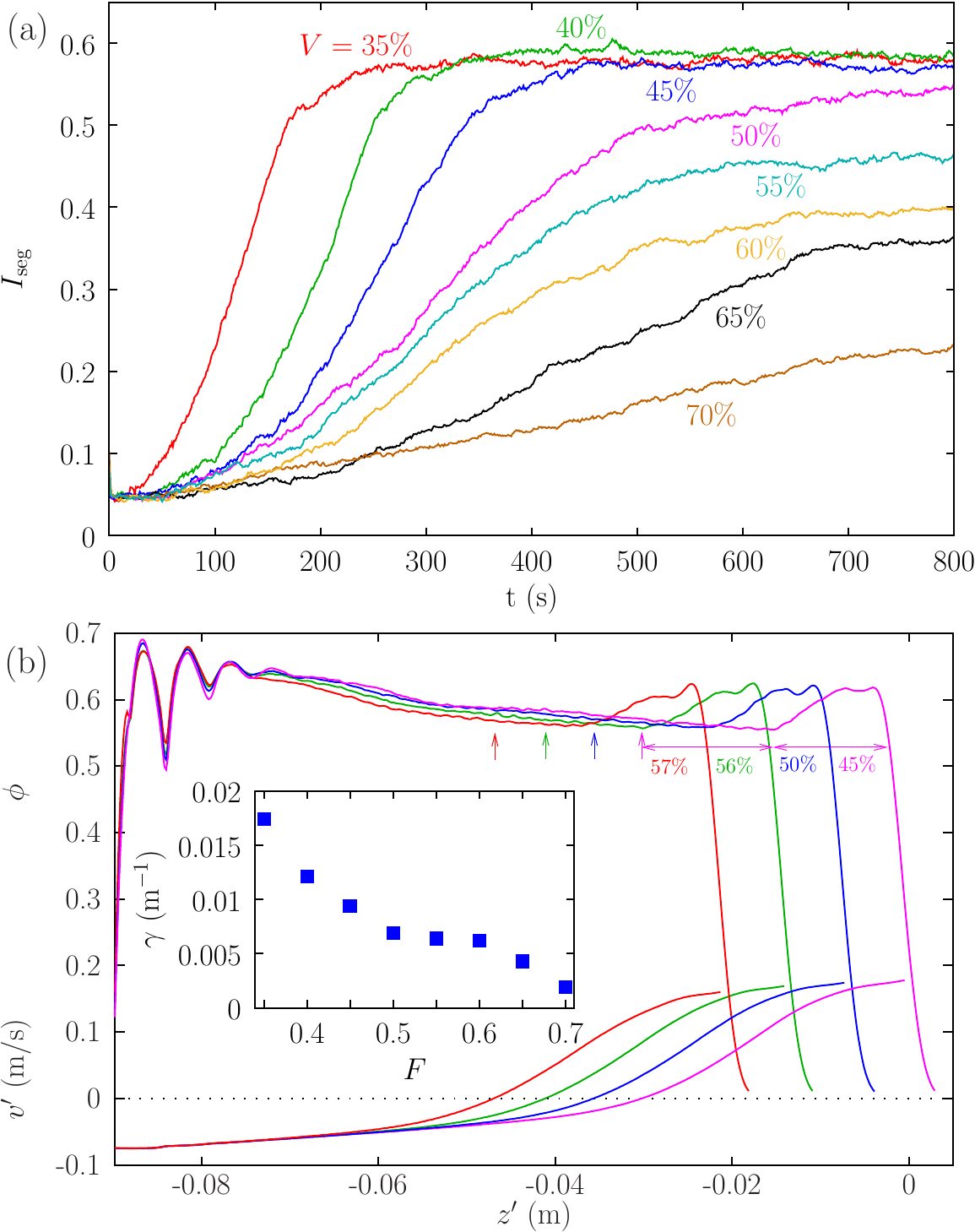}
\caption{(a) Time evolution of $I_\mathrm{seg}$ for fill levels $0.35 \leq V \leq 0.7$ with $L=100d$, $D=90d$, $\Delta f_l=0.1$, $\omega=15$~rpm, $d_l/d=3$ ($d=2$~mm), and $F_l=F_s=0.5$. See Fig.~S13 in the SM for results on a semi-log scale. (b) Profiles of $\phi$ at ($y'=0$) with $L=50d$ to prevent axial segregation. Vertical arrows indicate the location of the bottom of the flowing zone, corresponding to $v'=0$. 
Percentages indicate the fraction of the dense layer thickness compared to the flowing layer thickness where the thickness of the excess density zone is based on $\phi \geq$~min$(\phi)$ (see purple double-ended arrows for the case $V=50\%$).
Inset: growth rate as a function of the fill level. 
}
\label{d3fill}
\end{figure}

The volume fill level in the tumbler, $V$, 
also affects axial segregation. Results to this point have been obtained for a half-filled tumbler, $V=0.5$. Figure \ref{d3fill}(a) shows $I_\mathrm{seg}$ for $0.35 \leq V \leq 0.7$. Axial segregation is faster for lower fill levels, as would be expected simply because in a 35\% filled tumbler, particles pass more frequently through the flowing zone over a given amount of time than in a 50\% filled tumbler. The growth rate based on an exponential fit decreases from $\gamma=0.017$~s$^{-1}$ for $V=0.35$ to $\gamma=0.002$~s$^{-1}$ for $V=0.7$, as shown in the inset of Fig.~\ref{d3fill}(b).  But even when time is rescaled by assuming that the number of passes of particles through the flowing zone, $n$, is inversely proportional to $V$ (that is, $n \sim 1/V$) and plotting $I_{\rm seg}$ versus $n$, the $V=0.35$ case still segregates axially more rapidly by a factor of 2.

Figure~\ref{d3fill}(b) provides $\phi$ profiles and the associated streamwise velocity profiles for different fill levels. 
The bumps associated with the higher-density mixed layer are similar in height and width for all fill levels, but the bump becomes slightly wider with respect to the total thickness of the flowing layer, measured from the bottom of the flowing layer (where $v'=0$) to the surface (where $\phi = 0.3$), as $V$ decreases. While it is possible that this plays a role in the increased growth rate at small $V$, it seems unlikely that this effect alone is the cause. Other mechanisms, perhaps related to the geometry of the tumbler and particle bed, likely also play roles.

\section{Conclusions}

This paper has examined the axial segregation of size-bidisperse granular mixtures in horizontal rotating tumblers in the context of the granular Rayleigh-Taylor (RT) instability, a relatively recently discovered phenomenon that can occur in segregating bidisperse granular materials as they flow in a gravitational field~\cite{DOrtonaThomas20,DOrtonaMartinand25,DOrtonaLueptow26}. Key to the granular RT instability is that the density excess in an upper layer that drives the instability arises from a unique property of granular materials: a mixture of particles of different sizes packs more densely than monodisperse particles. We recently showed that it is the RT instability that drives axial segregation in a rotating tumbler~\cite{DOrtonaLueptow24}, finally providing the correct physical explanation for this phenomenon more than eight decades after it was first observed~\cite{Oyama40}. Here we analyze {\blue the underlying RT instability and how it drives the development of} the axial segregation phenomenon in greater detail to complement our earlier work first describing the mechanism~\cite{DOrtonaLueptow24}.

After demonstrating how segregation leads to a mixed-particle layer with a density excess over the underlying monodisperse layer of small particles, we show how the resulting RT instability {\blue amplifies with distance along the flowing layer (Figs.~\ref{colorinit} and~\ref{colorstatio}) and through repeated passes of particles through the flowing layer to eventually generate} streamwise rolls {\blue (Fig.~\ref{dispmap}) that then} amplify the segregated bands of large and small particles. {\blue Although we previously found that} axial segregation is sensitive to the particle size ratio~\cite{DOrtonaLueptow24}, similar to size bidisperse chute flow~\cite{DOrtonaLueptow26}, {\blue here we consider even smaller particle size ratios to further demonstrate how the size ratio affects particle packing (Fig.~\ref{volfracb}) showing that there is a slight tendency for segregation even for size ratios as small as $d_l/d=1.15$ (Fig.~\ref{seuildl})}. We confirm that it is indeed the packing density that drives the RT instability by slightly adjusting the density of the two particle species {\blue (Fig.~\ref{seuildens})}. We then use the degree of axial segregation and its growth rate to further explore how the interplay between particle size and the tumbler diameter affects the evolution and extent of axial segregation. {\blue The tumbler diameter must be large enough compared to the particle size, $D/d \gtrsim 70$ (Figs.~\ref{Isegvsd},~\ref{d2wxxxtime}, and~\ref{growthratevsD}), for axial bands to form. Similar} fractions of large and small particles ($F_l \approx F_s$) optimize axial segregation {\blue (Fig.~\ref{isegvsfs}) because adequate volumes of both species are available for bands to form. Finally, smaller tumbler fill levels tend to segregate axially faster than larger fill levels (Fig.~\ref{d3fill}).}

Nearly all of these results can be explained in terms of the granular RT instability, further verifying our conclusion that, despite many alternative mechanisms being proposed in the past, the appearance of a high-density mixed-particle layer above low-density layer of monodisperse small particles initiates the granular RT instability and subsequent axial segregation that leads to axial band formation. Questions remain. The faster evolution of axial segregation observed at lower fill levels is not yet fully understood. Particle properties such as the friction coefficient may also influence axial segregation by altering the material's flowability, and endwall effects could further accelerate band formation. {\blue Preliminary results show that reducing the friction coefficient from $\mu=0.7$ to 0.2 induces an increase of the growth rate by a factor of 2. A similar phenomenon was observed on the incline geometry \cite{DOrtonaLueptow26}, suggesting that the mechanism is likely not connected to the geometry of the tumbler but is linked to flowing properties of the granular material.} {\blue Preliminary simulations also suggest dependence of band formation on the Froude number, although the mechanism behind this dependence is still under investigation.} In addition, band merging and coarsening, which occurs experimentally for many more tumbler rotations than we can feasibly simulate, remains unexplained. Despite these open issues, the present work broadens our understanding of how the granular RT instability drives axial segregation of size-bidisperse granular materials in rotating tumblers.

\section*{Acknowledgments}
Centre de Calcul Intensif d'Aix-Marseille University is acknowledged for granting access to its high performance computing resources. This project was provided with computing HPC and storage resources by GENCI at TGCC thanks to the grant AD012B12432R3 on the supercomputer Joliot Curie's ROME partition.


\begin{thebibliography}{99}
\bibitem{OttinoKhakhar00}J. M. Ottino and D. V. Khakhar, Annu. Rev. Fluid
Mech. {\bf 32}, 55 (2000). 
\bibitem{MeierLueptow07}S. W. Meier, R. M. Lueptow, and J. M. Ottino, Adv. Phys. {\bf 56}, 757 (2007).
\bibitem{Ottino06} J. M. Ottino, Chem. Eng. Sci. {\bf 61}, 4165 (2006).
\bibitem{ClementRajchenbach95}E. Clement, J. Rajchenbach, and J. Duran,
Europhys. Lett. {\bf 30}, 7 (1995).
\bibitem{CantelaubeBideau95}F. Cantelaube and D. Bideau, Europhys. Lett. {\bf 30}, 133 (1995).
\bibitem{DuryRistow1997}C. M. Dury and G. H. Ristow, J. Phys. I France {\bf 7}, 737 (1997).
\bibitem{KhakharMcCarthy97}D. V. Khakhar, J. J. McCarthy, and J. M. Ottino, Phys. Fluids {\bf 9}, 3600 (1997).
\bibitem{Oyama40}Y. Oyama, Sci. Pap. Phys. Chem. Res. (Jpn.) {\bf 37}, 17 (1940).
\bibitem{DonaldRoseman62}M. B. Donald and B. Roseman, Br. Chem. Eng. {\bf 7}, 749 (1962). 
\bibitem{DasguptaKhakhar91} S. Das Gupta, D. V. Khakhar, and S. K.
Bhatia, Chem. Eng. Sci. 46, 1513 (1991).
\bibitem{Nakagawa94}M. Nakagawa, Chem. Eng. Sci. {\bf 49}, 2540 (1994).
\bibitem{ZikLevine94}O. Zik, D. Levine, S. G. Lipson, S. Shtrikman, and J. Stavans, Phys. Rev. Lett. {\bf 73}, 644 (1994).
\bibitem{HillKakalios94}K. M. Hill and J. Kakalios, Phys. Rev. E {\bf 49}, R3610 (1994).
\bibitem{HillKakalios95}K. M. Hill and J. Kakalios, Phys. Rev. E {\bf 52}, 4393 (1995).
\bibitem{Rapaport02}D. C. Rapaport, Phys. Rev. E {\bf 65}, 061306 (2002).
\bibitem{ChooMorris98}K. Choo, M. W. Baker, T. C. A. Molteno and S. W. Morris, Phys. Rev. E {\bf 58}, 6115 (1998).
\bibitem{Levine99}D. Levine, Chaos {\bf 9}, 573 (1999).
\bibitem{AlexanderMuzzio04}A. Alexander, F. J. Muzzio, and T. Shinbrot, Gran. Matter {\bf 5}, 171 (2004).
\bibitem{CharlesMorris06}C. R. J. Charles, Z. S. Khan, and S. W. Morris, Gran. Matter {\bf 8}, 1 (2006). 
\bibitem{FiedorOttino06}S. J. Fiedor, P. Umbanhowar, and J. M. Ottino, Phys. Rev. E {\bf 73}, 041303 (2006).
\bibitem{JuarezLueptow08}G. Juarez, J. M. Ottino, and R. M. Lueptow, Phys. Rev. E {\bf 78}, 031306 (2008).
\bibitem{JuarezOttino10}G. Juarez, R. M. Lueptow, and J. M. Ottino, Phys. Rev. E {\bf 81}, 012301 (2010).
\bibitem{ChooMorris97}K. Choo, T. C. A. Molteno and S. W. Morris, Phys. Rev. Lett. {\bf 79}, 2975 (1997).
\bibitem{Frette97} V. Frette and J. Stavans, Phys. Rev. E {\bf 56}, 6981 (1997).
\bibitem{HillCaprihan97}K. M. Hill, A. Caprihan, and J. Kakalios, Phys.
Rev. E 56, 4386 (1997).
\bibitem{FiedorOttino03}S. J. Fiedor and J. M. Ottino, Phys. Rev. Lett. {\bf 91}, 244301 (2003). 

\bibitem{DOrtonaThomas16} U. D'Ortona, N. Thomas, and R. M. Lueptow, Phys. Rev. E 93 022906 (2016).
\bibitem{DOrtonaThomas17} U. D'Ortona, N. Thomas, and R. M. Lueptow. EPJ web of conferences {\bf 140}, 03011 (2017).
\bibitem{ChenLochman09} P. Chen, B. J. Lochman, J. M. Ottino and R. M. Lueptow, Phys. Rev. Lett. 102, 148001 (2009).
\bibitem{JainKhakhar01}N. Jain, D. V. Khakhar, R. M. Lueptow, and J. M. Ottino, Phys. Rev. Lett. {\bf 86}, 3771 (2001).
\bibitem{Arndt05}T. Arndt, T. Siegmann-Hegerfeld, S. J. Fiedor, J. M. Ottino, and R. M. Lueptow, Phys. Rev. E {\bf 71}, 011306 (2005).
\bibitem{KhanMorris2004} Z. S. Khan, W. A. Tokaruk, and S. W. Morris, Europhys. Lett {\bf 66}:212-218 (2004).
\bibitem{TaberletNewey06}N. Taberlet, M. Newey, P. Richard, and W. Losert, J. Stat. Mech. P07013 (2006).
\bibitem{ChenLueptow11}P. Chen, J. M. Ottino, and R. M. Lueptow, New J. Phys. {\bf 13}, 055021 (2011).
\bibitem{Aranson1999a} I. S. Aranson and L. S. Tsimring, Phys. Rev. Lett. {\bf 82}, 4643 (1999).
\bibitem{Aranson1999b} I. S. Aranson, L. S. Tsimring, and V. M. Vinokur, Phys. Rev. E {\bf 60}, 1975 (1999).
\bibitem{DOrtonaLueptow24} U. D'Ortona, R. M. Lueptow, and N. Thomas, Phys. Rev. Research {\bf 6}, L032038  (2024).
\bibitem{DOrtonaThomas20} U. D'Ortona and N. Thomas, Phys. Rev. Lett. {\bf 124}, 178001 (2020).
\bibitem{DOrtonaLueptow26} U. D'Ortona, R. M. Lueptow, and N. Thomas, Phys. Rev. E {\bf 112}, 025404 (2026).
\bibitem{DOrtonaMartinand25} U. D'Ortona, D. Martinand, and N. Thomas, Phys. Rev. E {\bf 112}, 015421 (2025).
\bibitem{Pilliteri20} S. Pillitteri, E. Opsomer, G. Lumay, and N. Vandewalle, Soft Matter, {\bf 16}, 9094 (2020).
\bibitem{CundallStrack79} P. A. Cundall and O. D. L. Strack, Geotechnique {\bf 29}, 47 (1979).
\bibitem{Kloss2012} C. Kloss, C. Goniva, A. Hager, S. Amberger, and S. Pirker, Prog. Comput. Fluid Dyn. {\bf 12}, 140 (2012). 
\bibitem{SchaferDippel96}J. Sch\"afer, S. Dippel, and D. E. Wolf, J. Phys. I (France) {\bf 6}, 5 (1996).
\bibitem{DrakeShreve86}T. G. Drake and R. L. Shreve, J. Rheol. {\bf 30}, 981 (1986).
\bibitem{FoersterLouge94}S. F. Foerster, M. Y. Louge, H. Chang, and K. Allia, Phys. Fluids {\bf 6}, 1108 (1994).
\bibitem{ZamanDOrtona13}Z. Zaman, U. D'Ortona, P. B. Umbanhowar, J. M. Ottino, and R. M. Lueptow, Phys. Rev. E {\bf 88}, 012208 (2013).
\bibitem{SM} Supplementary Material, href will be inserted by the Editor. Supplementary figures; Section showing that a poor time resolution induces a high axial segregation; Section showing the link between the segregation index and the amplitude of the RTI perturbation.
\bibitem{Chandrasekhar61} S. Chandrasekhar, {\sl Hydrodynamic and Hydromagnetic Stability} (Oxford University Press, London, 1961).
\bibitem{Charru} F. Charru, {\sl Hydrodynamic Instabilities} (Cambridge University Press, Cambridge, 2011)
\bibitem{Thomas00} N. Thomas, Phys. Rev. E 62, 961 (2000).
\bibitem{DOrtonaThomas18} N. Thomas and U. D'Ortona, Phys. Rev. E 97, 022903 (2018).
\bibitem{Dodds80} J. A. Dodds, J. Colloid. Interface Sci. {\bf 77}, 317 (1980).
\bibitem{Vyas24} D. R. Vyas, S. Gao, P. B. Umbanhowar, J. M. Ottino, R. M. Lueptow, AIChE J. {\bf 70}, e18499 (2024).
\bibitem{Selig65}F. Selig, Geophysics {\bf 4}, 633 (1965).
\bibitem{Ramberg68} H. Ramberg, Phys.  Planet. Interiors {\bf 1}, 63 (1968).
\end{thebibliography}
\end{document}